\documentclass[a4paper,fleqn]{cas-dc}
\usepackage{graphicx}
\usepackage{float}
\usepackage{subfig}
\usepackage{amsmath,amsfonts,amssymb}
\usepackage{multirow}
\usepackage{graphicx}
\usepackage{booktabs,makecell,multirow,array,colortbl,dcolumn}
\RequirePackage[svgnames,dvipsnames]{xcolor}
\usepackage{appendix}
\usepackage{threeparttable}
\usepackage{pifont}
\usepackage{ulem}
\usepackage{makecell}
\usepackage{placeins}

\usepackage[authoryear,round,sort]{natbib}

\def\tsc#1{\csdef{#1}{\textsc{\lowercase{#1}}\xspace}}
\tsc{WGM}
\tsc{QE}

\begin{document}
\let\WriteBookmarks\relax
\def\floatpagepagefraction{1}
\def\textpagefraction{.001}

\shorttitle{Improve Retinal Artery/Vein Classification via Channel Coupling}    

\shortauthors{Shuang Zeng et~al.}  

\title [mode = title]{Improve Retinal Artery/Vein Classification via Channel Coupling}

\author[1,2,3,4]{Shuang Zeng}[orcid=0009-0004-1936-3802]\ead{stevezs@pku.edu.cn}
\author[1]{Chee Hong Lee}\ead{2100098602@stu.pku.edu.cn}
\author[1,2,3]{Kaiwen Li}\ead{kaiwenli325@gmail.com}
\author[1]{Boxu Xie}\ead{2310117138@stu.pku.edu.cn}
\author[1,2,3]{Ourui Fu}\ead{orfu@stu.pku.edu.cn}
\author[1,2,3]{Hangzhou He}\ead{zhuang@stu.pku.edu.cn}
\author[1,2,3]{Lei Zhu}\cormark[1]\ead{zhulei@pku.edu.cn}
\author[1,2,3]{Yanye Lu}\cormark[1]\ead{yanye.lu@pku.edu.cn}
\author[1]{Fangxiao Cheng}\cormark[1]\ead{chengfangxiao@bjmu.edu.cn}

\address[1]{Institute of Medical Technology, 
Peking University Health Science Center, Peking University, Beijing, China}
\address[2]{Department of Biomedical Engineering, Peking University, Beijing, China}
\address[3]{National Biomedical Imaging Center, Peking University, Beijing, China}
\address[4]{Wallace H. Coulter Department of Biomedical Engineering, Georgia Institute of Technology and Emory University, Atlanta, GA, USA}
\cortext[1]{Corresponding authors: Lei Zhu, Yanye Lu and Fangxiao Cheng at Institute of Medical Technology, 
Peking University Health Science Center, Peking University, Beijing, China.}

\begin{abstract}
    Retinal vessel segmentation plays a vital role in analyzing fundus images for the diagnosis of systemic and ocular diseases. Building on this, classifying segmented vessels into arteries and veins (A/V) further enables the extraction of clinically relevant features such as vessel width, diameter and tortuosity, which are essential for detecting conditions like diabetic and hypertensive retinopathy. However, manual segmentation and classification are time-consuming, costly and inconsistent. With the advancement of Convolutional Neural Networks, several automated methods have been proposed to address this challenge, but there are still some issues. For example, the existing methods all treat artery, vein and overall vessel segmentation as three separate binary tasks, neglecting the intrinsic coupling relationships between these anatomical structures. Considering artery and vein structures are subsets of the overall retinal vessel map and should naturally exhibit prediction consistency with it, we design a novel loss named Channel-Coupled Vessel Consistency Loss to enforce the coherence and consistency between vessel, artery and vein predictions, avoiding biasing the network toward three simple binary segmentation tasks. Moreover, we also introduce a regularization term named intra-image pixel-level contrastive loss to extract more discriminative feature-level fine-grained representations for accurate retinal A/V classification. SOTA results have been achieved across three public A/V classification datasets including RITE, LES-AV and HRF. Our code will be available upon acceptance. 
\end{abstract}










\maketitle
\section{Introduction}
\label{sec1}

\begin{figure}[t]
\centering
\includegraphics[width=0.49\textwidth]{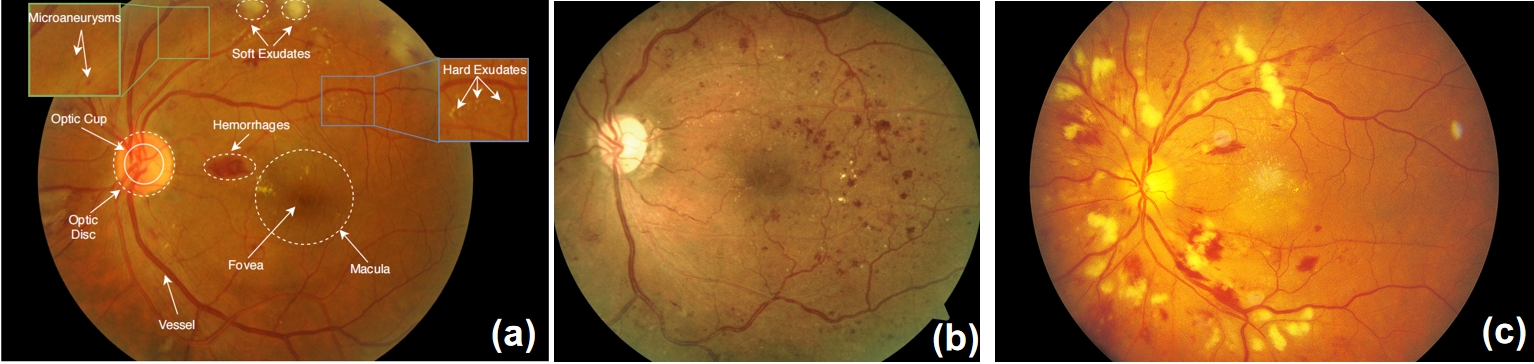}
\caption{(a) A fundus image from IDRiD dataset illustrating important biomarkers and lesions. (b) An example of Diabetic Retinopathy fundus image. (c) An example of Hypertensive Retinopathy fundus image.}
\label{background}
\end{figure}

The morphological characteristics of retinal blood vessels (BV) in Figure \ref{background}(a), such as their caliber and geometric arrangement, serve as critical biomarkers for the diagnosis and monitoring of a range of systemic and ocular conditions. For example, Diabetic Retinopathy (DR), a common complication of diabetes, results from prolonged high blood glucose that lead to vessel leakage and swelling \citep{smart2015study}, as illustrated in Figure \ref{background}(b). Likewise, Hypertensive Retinopathy (HR), caused by elevated blood pressure, induces structural changes in retinal vasculature, such as vessel narrowing and tortuosity \citep{ding2014retinal}, as shown in Figure \ref{background}(c). These vascular alterations can be effectively assessed by trained ophthalmologists through the analysis of color fundus images captured via retinography --- a non-invasive, cost-effective imaging modality. Owing to its accessibility and non-invasiveness, retinography is extensively utilized in clinical diagnostics, epidemiological studies, and large-scale screening programs. 

\begin{figure}[t]
\centering
\includegraphics[width=0.49\textwidth]{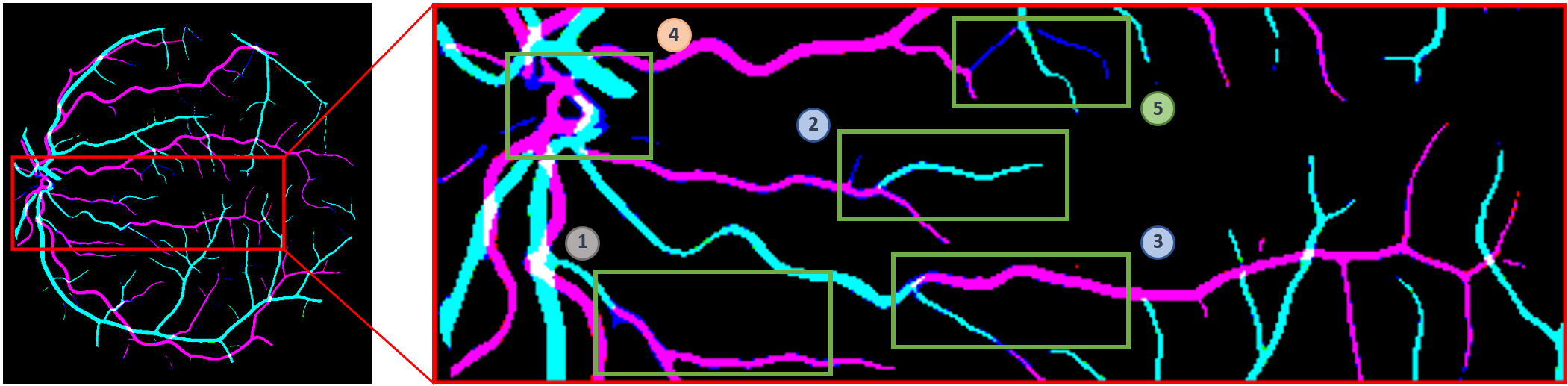}
\caption{Examples of common {\itshape manifest classification errors} produced by the SOTA FCNN-based method: RRWNet \citep{morano2024rrwnet}, Magenta pixels indicate \textcolor{Magenta}{arteries}; cyan pixels indicate \textcolor{cyan}{veins}; blue pixels indicate \textcolor{blue}{uncertain} vessel regions; white pixels indicate \textcolor{gray}{crossing} areas. (1) While most of the vessel is classified as \textcolor{cyan}{vein}, the model misclassifies the distal part as \textcolor{Magenta} {artery}. (2-3) The presence of vascular bifurcations can occasionally hinder the model's ability to accurately differentiate between artery and vein. (4) The model often misclassifies vessels in crossing areas, especially in optic disc. (5) Micro vessels cannot be accurately classified. These {\itshape manifest classification errors} are easily detected by a human observer because they are inconsistent with the overall structure of the vascular tree.}
\label{AV error}
\end{figure}

A detailed evaluation of the retinal vasculature necessitates the segmentation of blood vessels and their classification into arteries and veins (A/V). This yields separate A/V segmentation maps in Figure \ref{AV error} left, which supports the extraction of various diagnostically relevant features such as vessel width, diameter, and tortuosity. However, manual segmentation and classification are time-consuming, costly, and susceptible to inter-observer variability, thereby limiting reproducibility and diagnostic consistency. To overcome these challenges, numerous automated methods have been proposed to perform simultaneous vessel segmentation and A/V classification \citep{mookiah2021review}. 

With the advancements in machine learning and computer vision, deep learning frameworks have become competitive in A/V classification and provided detailed vascular features from retinal images, aiding clinicians in diagnosing and treating various eye diseases. Current state-of-the-art methods for A/V classification predominantly rely on Fully Convolutional Neural Networks (FCNNs) \citep{long2015fully}, which have demonstrated strong performance across various medical image segmentation tasks. Most approaches \citep{galdran2022state, hemelings2019artery, hu2024semi, karlsson2022artery} formulate the task as a four-class semantic segmentation problem, assigning each pixel to one of the following categories: background, artery, vein or crossing ({\itshape i.e.} regions where arteries and veins intersect). Additionally, some methods \citep{galdran2019uncertainty} incorporate an ``uncertain" class to account for pixels presenting ambiguous characteristics. In contrast, some recent approaches \citep{chen2022tw, morano2021simultaneous, morano2024rrwnet} reformulate the problem as a multi-label segmentation task, enabling the network to independently predict the presence of arteries, veins and blood vessels ({\itshape i.e.} both arteries and veins) by allowing pixels to be assigned to multiple classes simultaneously.    

Despite their architectural and formulation improvement, FCNN-based methods consistently encounter a major challenge: {\itshape manifest classification errors}. These errors often appear as locally inconsistent or contradictory predictions within otherwise correctly segmented vessels, undermining the anatomical plausibility of the results shown in Figure \ref{AV error}. Such errors arise from the propensity of FCNN-based models to classify vessels based on local characteristics of the input image, overlooking the global structural context (such as topology, connectivity, bifurcation) of the vascular tree. 
To alleviate these issues, some methods employ {\itshape ad hoc} post-processing techniques. Specifically, AV-casNet \citep{AV-casNet} employs a two-stage framework in which a CNN module first produces an initial segmentation, followed by a cascaded graph neural network (GNN) module that refines vessel connectivity. TW-GAN \citep{chen2022tw} proposes an end-to-end topology (including a topology-ranking discriminator and a topology-preserving regularization module to improve vascular connectivity) and a width-aware network for A/V classification. RRWNet \citep{morano2024rrwnet} proposes an end-to-end deep learning framework that recursively refines semantic segmentation maps to correct classification errors and enhance topological consistency. 

These methods have achieved promising results on A/V classification by introducing multi-stage framework, integrating specific vessel information or designing recursive refinement subnetwork. Nevertheless, there are still several issues to be solved: 
(1) All the existing methods treat artery, vein and overall vessel segmentation as three separate binary tasks, optimized independently using losses like Binary Cross-Entropy (BCE) loss. However, this strategy neglects the intrinsic coupling relationships between these anatomical structures. Specifically, artery and vein structures are subsets of the overall retinal vessel map and should naturally exhibit prediction consistency with it. Ignoring this interdependence can lead to inconsistencies between A/V map and the vascular topology.
(2) The goal of retinal A/V classification is to assign a class label (artery or vein) to each pixel in a fundus image, with an emphasis on capturing intra-image differences.  
Therefore, it is vital for models to extract more discriminative and fine-grained pixel-level features. However, most existing approaches prioritize minimizing the discrepancy between final predictions and labels through various loss functions, while underutilizing rich feature representations extracted by the encoder. This often leads to suboptimal performance in distinguishing arteries from veins, especially in challenging regions like vessel crossings or peripheral branches.


\begin{figure}[t]
\centering
\includegraphics[width=0.5\textwidth]{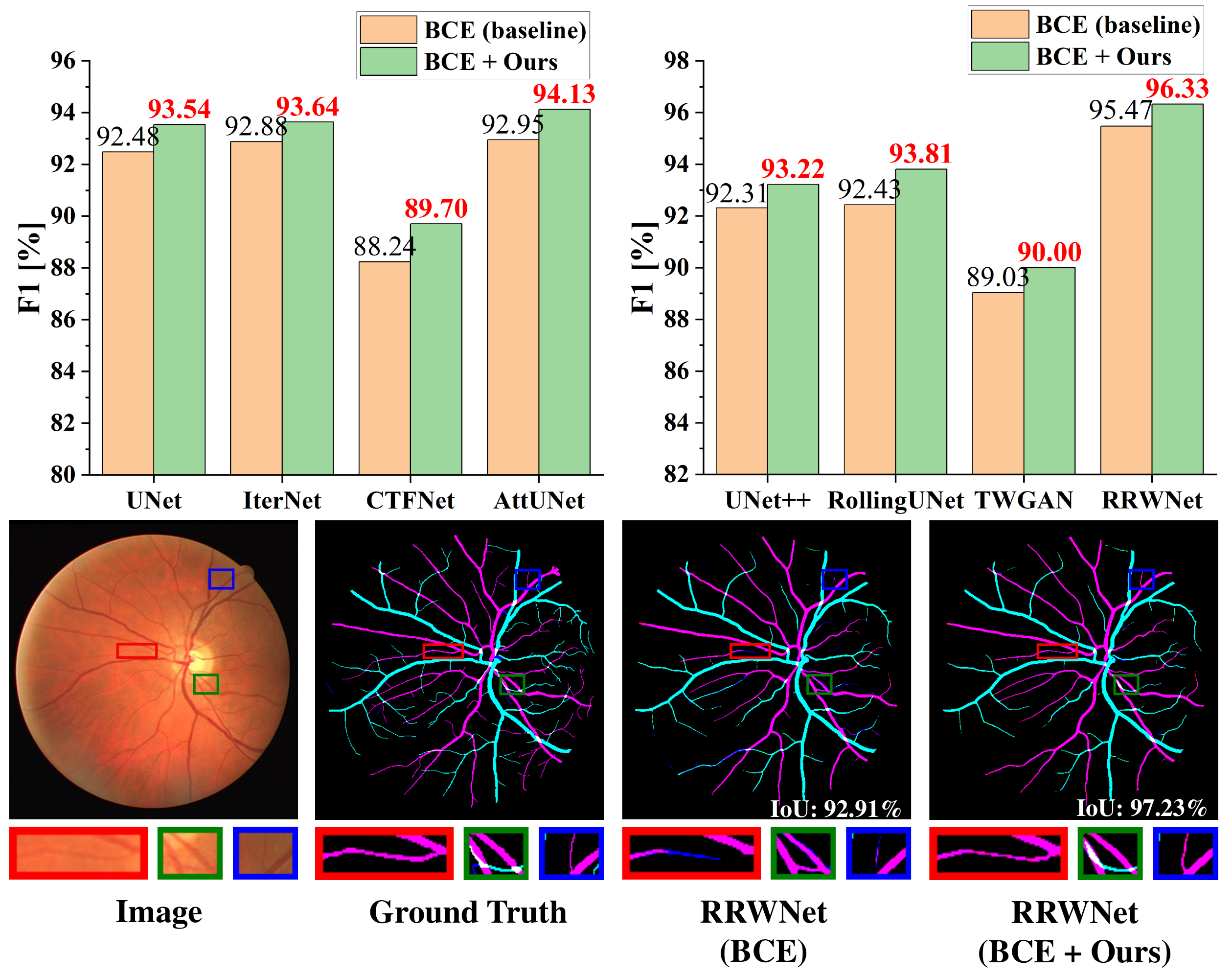}
\caption{We demonstrate the superiority of our proposed method on RITE from both quantitative and qualitative perspectives: (1) Our designed loss function yields promising improvements across all 8 vessel segmentation backbones. (2) Visualization results show that networks trained with our proposed loss can produce more accurate A/V segmentation maps, especially at bifurcation vessels or distal micro vessels.}
\label{backbone_ablation}
\end{figure}

To address the above issues, in this work, we propose a novel loss function named \textbf{C}hannel-\textbf{C}oupled Vessel \textbf{C}onsistency ($C^3$) Loss. $C^3$ loss addresses the lack of inter-channel consistency in previous methods, which treat artery, vein and vessel predictions as independent tasks. By constructing a fused prediction map that considers the anatomical relationships among these three channels, $C^3$ loss enforces consistency and coherence across artery, vein and vessel channels. Furthermore, we also introduce the intra-image pixel-level contrastive loss \citep{SuperCL} as a regularization term to enable the network to capture more discriminative feature-level fine-grained representations by treating pixels within the same superpixel cluster as positive pairs and those from different clusters as negatives. As shown in Figure \ref{backbone_ablation}, our proposed method achieves promising performance from both quantitative and qualitative perspectives. To sum up, the main contributions of this paper are as follows:

\begin{itemize}
\item {A novel loss named Channel-Coupled Vessel Consistency Loss is designed to enforce the coherence and consistency between vessel, artery and vein predictions, avoiding biasing the network toward three simple binary segmentation tasks.}

\item {In order to make the network capture more discriminative feature-level fine-grained representations for accurate retinal A/V classification, a regularization term named intra-image pixel-level contrastive loss is introduced by leveraging the structural coherence of superpixels to guide contrastive learning in an unsupervised manner.}

\item {State-of-the-art results have been achieved across three public A/V classification datasets including RITE, LES-AV and HRF. Comprehensive experiments and ablation studies are also conducted to verify the generalization ability and the effectiveness of the losses.}
\end{itemize}

\section{Related Work}

\label{sec2}
\subsection{Retinal Vessel Segmentation}

Early methods for retinal vessel segmentation predominantly employ unsupervised techniques grounded in classical image processing operations, including filtering, thresholding, mathematical morphology, and edge detection \citep{singh2015local, zana2001segmentation, oliveira2016unsupervised}. Although these approaches offer initial solutions for vessel delineation, their effectiveness is constrained by the dependence on manually designed features and rigid rule-based frameworks.
With the advent of deep learning, more advanced and accurate segmentation techniques \citep{ronneberger2015u, li2020iternet, wang2020ctf, oktay2018attention, zhou2018unet++, liu2024rolling, zeng2025novel} emerge for retinal vessel segmentation. UNet \citep{ronneberger2015u} distinguishes itself as a milestone through its effective encoder-decoder architecture and skip connection, which enable precise delineation of anatomical structures. As a result, numerous UNet variants have been developed for retinal vessel segmentation tasks. For instance, IterNet \citep{li2020iternet} utilizes multiple iterations of mini-UNet to recover vessel details, and CTFNet \citep{wang2020ctf} adopts a coarse-to-fine supervision strategy to progressively refine segmentation outcomes. AttUNet \citep{oktay2018attention} integrates attention gates into skip connections to suppress irrelevant feature responses and enhance predictive accuracy. UNet++ \citep{zhou2018unet++} proposes a nested architecture with dense skip connections to improve feature fusion and segmentation precision. Moreover, RollingUNet \citep{liu2024rolling} combines MLP with UNet to efficiently fuse local features and long-range dependencies.

Moreover, some researchers also design specific loss functions to extract the structural context (such as topology, connectivity, bifurcation) of the vascular tree to enhance vessel segmentation. In detail, Connection Sensitive Loss \citep{li2019connection} proposes a connection sensitive loss to enhance the continuity of segmented vessels by penalizing disconnected predictions, thereby preserving vessel connectivity. TopoLoss \citep{hu2019topology} designs a continuous-valued loss that enforced the predicted segmentation to share the same topology as the ground truth, measured by matching Betti numbers. Flow-based Loss \citep{jena2021self} proposes a self-supervised method, using tube-like structure properties, such as connectivity, consistent profiles, and bifurcations as inductive biases to guide learning. Supervoxel-based Loss \citep{grim2025efficient} extends the concept of simple voxels to supervoxels and introduces a differentiable loss function that guides neural networks to minimize split and merge errors by preserving structural connectivity.

\subsection{A/V classification}

Until recently, A/V classification is typically approached as a two-step progress. In this paradigm, A/V classification is applied exclusively to pixels previously identified as blood vessels (BV) through a separate vessel segmentation algorithm \citep{estrada2015retinal, welikala2017automated}. Although these methods demonstrate reasonable performance, they are limited by the quality of the initial vessel segmentation. To address this issue, more recent research has focused on joint classification of retinal vessels. These efforts typically formulate the task as a multi-label target classes ({\itshape e.g.}, artery, vein, blood vessel) \citep{AV-casNet, chen2022tw, morano2024rrwnet}. This approach offers the advantage of generating continuous and topologically coherent segmentation maps, particularly at vessel crossings, which can be simultaneously attributed to both artery and vein classes. Specifically, AV-casNet \citep{AV-casNet} introduces a two-stage architecture, wherein an initial vessel segmentation is generated by a CNN, and subsequently refined through a cascaded GNN module designed to enhance vessel connectivity. In contrast, TW-GAN \citep{chen2022tw} presents an end-to-end framework that incorporates a topology-aware design, featuring a topology-ranking discriminator and a topology-preserving regularization component, both aimed at improving vascular structure continuity and preserving vessel width for effective A/V classification. Meanwhile, RRWNet \citep{RRWNet} proposes an end-to-end deep learning approach that recursively refines the semantic segmentation output, effectively correcting classification errors and reinforcing topological consistency throughout the vascular network. Furthermore, the existing methods for A/V classification primarily focus on modifications to network architectures, without considering how to leverage the intrinsic relationships between artery, vein and blood vessel from the perspective of the loss function design. 

\subsection{Superpixel Segmentation}
Superpixel segmentation aims to group perceptually similar neighboring pixels into compact and meaningful regions, serving as pre-processing step to reduce image complexity. Traditional methods are generally divided into clustering-based and graph-based approaches. Clustering-based methods, such as SLIC \citep{SLIC}, SNIC \citep{SNIC} and LSC \citep{LSC}, typically employ classical clustering techniques like k-means to compute the connectivity between the anchor pixels and its neighbors. Specifically, SLIC \citep{SLIC} improves efficiency by restricting the clustering to a local neighborhood. SNIC \citep{SNIC} further speeds up computation via a non-iterative clustering strategy that updates cluster centers and pixel labels simultaneously. LSC \citep{LSC} enhances clustering quality by approximating normalized cuts through weighted k-means. Graph-based methods, like FH \citep{FH} and ERS \citep{ERS}, constructed an undirected graph based on image features. FH \citep{FH} merges regions based on edge weights in a minimum spanning tree, while ERS \citep{ERS} maximizes entropy by incrementally adding edges to the graph. With the rise of deep learning, CNN-based superpixel methods have emerged. SEAL \citep{SEAL} introduces a segmentation-aware loss but lacks full differentiability. SSN \citep{SSN} builds a differentiable framework inspired by SLIC, though it relies on labeled supervision and iterative center updates. SuperpixelFCN \citep{SuperpixelFCN} simplifies label assignment via grid-based prediction, still under supervision. To overcome this, LNSNet \citep{LNSNet} proposes an unsupervised, lifelong clustering strategy to learn superpixels without manual labels.

\subsection{Contrastive Learning}

In recent years, Contrastive Learning (CL) \citep{chen2020simple, he2020momentum, GCL, PCL, MACL, SuperCL} has achieved notable success in learning discriminative representations from unlabeled data, substantially reducing the reliance on costly manual annotated data. The core idea of CL is to bring similar representations closer while pushing dissimilar ones apart by constructing positive and negative sample pairs. This paradigm has been widely used in self-supervised representation learning. For example, 
SimCLR \citep{chen2020simple} utilizes large batch sizes to ensure diverse negative pairs, while MoCo \citep{he2020momentum} adopts a momentum encoder and a queue-based dictionary for consistent feature comparison. In the medical domain, CL has been adapted to leverage domain-specific cues: GCL \citep{GCL} exploits structural consistency via partition-based strategies, and PCL \citep{PCL} incorporates spatial positional information to generate more meaningful contrastive pairs.

\begin{figure*}[t]
\centering
\includegraphics[width=\textwidth]{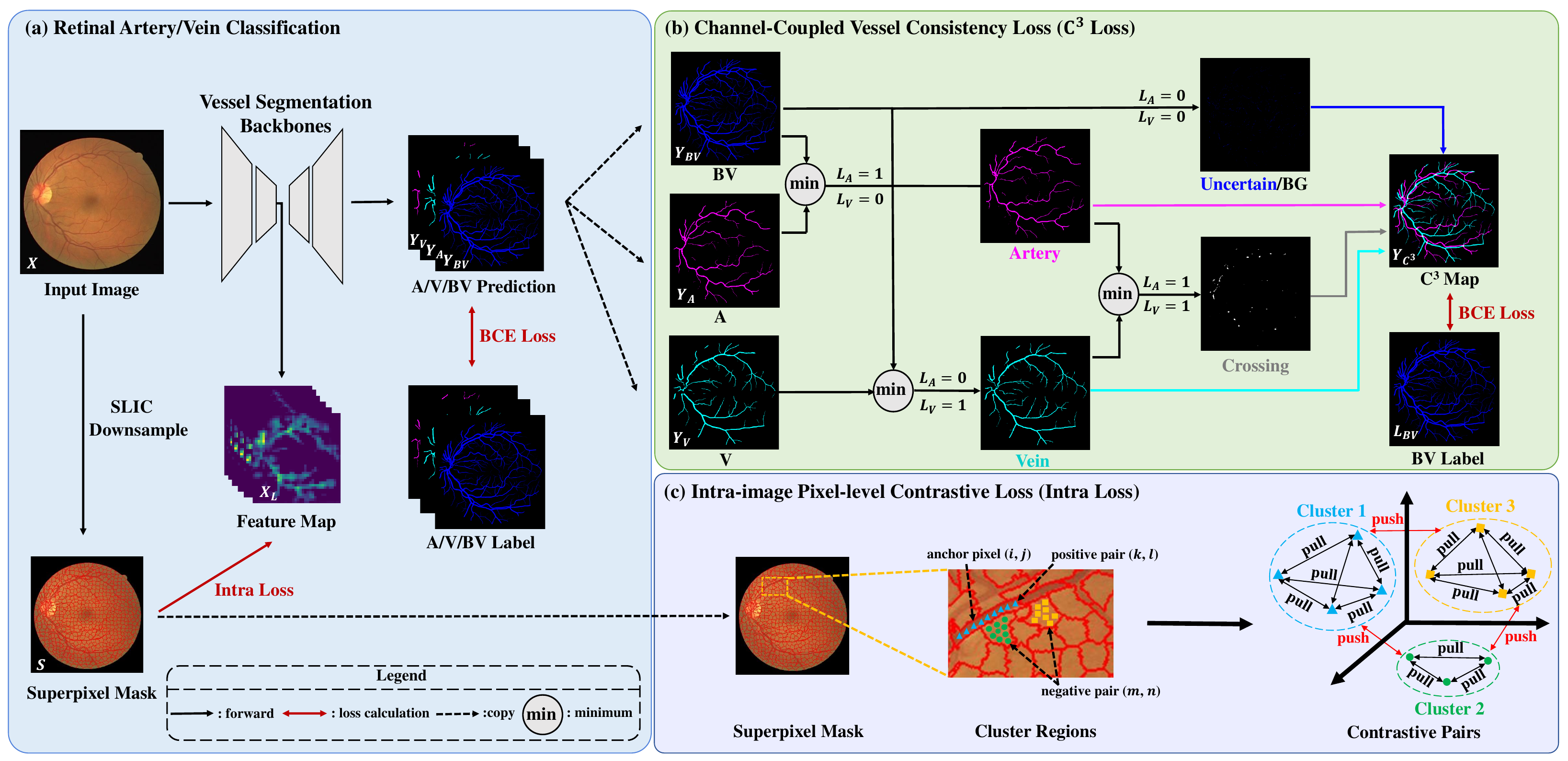}
\caption{Overview of our proposed method. (a) Illustration of retinal artery/vein classification pipeline. (b) Our proposed Channel-Coupled Vessel Consistency Loss ($C^3$ Loss): With the original A (artery), V (vein) and BV (blood vessel) prediction map output from the network, we can use the minimum operation to fuse the specialized knowledge of different classes, including Artery, Vein, Crossing, Uncertain blood vessel and Background region and get the modified $C^3$ map. Then we optimize the network by calculating the BCE loss between the $C^3$ map and BV label so as to enhance the coherence and consistency between vessel, artery adn vein predictions and avoid biasing the network toward these three simple binary segmentation tasks. (c) The introduced regularization term named Intra-image Pixel-level Contrastive Loss (Intra Loss).}
\label{network}
\end{figure*}

\section{Methodology}\label{sec3}
This section focuses on introducing the two losses, including the novel Channel-Coupled vessel Consistency loss ($\mathcal{L}_{C^3}$) and the regularization term named intra-image pixel-level contrastive loss ($\mathcal{L}_{intra}$). Firstly, a brief overview of retinal A/V classification is provided in Section \ref{sec3.1}. Then the designed $\mathcal{L}_{C^3}$ and introduced ($\mathcal{L}_{intra}$) will be discussed in Section \ref{sec3.2} and \ref{sec3.3}, respectively.

\subsection{Overview}\label{sec3.1}
The pipeline of our proposed method for retinal A/V classification is illustrated in Figure. \ref{network}(a). Given an input retinal fundus image $\boldsymbol{X}\in \mathbb{R}^{C \times H \times W}$, where $H \times W$ signifies the spatial resolution of the image and $C$ denotes the number of channels (3 for RGB retinography images and 1 for grayscale images), retinal A/V classification task aims to generate the corresponding pixel-wise classification map $\boldsymbol{Y} \in \mathbb{R}^{3 \times H \times W}$, which has three channels, corresponding to blood vessel (BV), artery (A) and vein (V). To achieve this purpose, the segmentation network needs an encoder $e(\cdot)$ to extract multi-level features and then a decoder $d(\cdot)$ is used to fuse features into the final segmentation map $\boldsymbol{Y}$ to recover image details: 
\begin{equation}
\boldsymbol{Y}=d(e(\boldsymbol{X}))=d\left(\left\{\boldsymbol{X}^1, \cdots, \boldsymbol{X}^L\right\}\right)
\end{equation}
where $\boldsymbol{X}^ \ell\in\mathbb{R}^{c_\ell \times {h}_{\ell} \times {w}_{\ell}}$ denotes the $\ell_{t h}$-level feature, $\ell \in \{1,\cdots,L\}$, $L$ denotes the number of encoder layers, $c_\ell$ denotes the channels of $\boldsymbol{X}^ \ell$, and $h_\ell \times w_\ell$ denotes the spatial size of $\boldsymbol{X}^ \ell$. 

To optimize this network, we utilize a baseline binary cross-entropy loss $\mathcal{L}_{BCE}$ along with our proposed Channel-Coupled vessel Consistency loss $\mathcal{L}_{C^3}$ and a regularization term named intra-image pixel-level contrastive loss $\mathcal{L}_{intra}$. Specifically, the final loss can be formulated as:
\begin{equation}
\mathcal{L}_{all}=\mathcal{L}_{BCE}+
\lambda_1 \times \mathcal{L}_{C^3} + \lambda_2 \times \mathcal{L}_{intra}
\end{equation}
where $\lambda_1, \lambda_2$ are the weighting coefficients. $\mathcal{L}_{BCE}$ is a fundamental baseline loss. $\mathcal{L}_{C^3}$ and $\mathcal{L}_{intra}$  will be discussed in Section \ref{sec3.2} and \ref{sec3.3}, respectively.

\subsection{Channel-Coupled Vessel Consistency Loss}\label{sec3.2}

In retinal A/V classification, the network outputs three prediction maps: the overall blood vessel $\boldsymbol{Y}_{BV}$, artery $\boldsymbol{Y}_A$ and vein $\boldsymbol{Y}_V$. Compared with the previous method \citep{RRWNet} which independently optimizes the three segmentation task with BCE loss, we propose a novel $C^3$ loss ($\mathcal{L}_{C^3}$) to enhance the coherence and consistency between the vessel, artery and vein predictions and avoid biasing the network toward these three simple binary segmentation tasks. 
Specifically, as shown in Figure. \ref{network}(c), 
we can get the $C^3$ map $\boldsymbol{Y}_{C^3}$ as follows:
\begin{equation}
\scalebox{0.8}{
$\begin{aligned}
\boldsymbol{Y}_{C^3} = \begin{cases}
min(\boldsymbol{{Y}}_{A}, \boldsymbol{{Y}}_{BV}), & \text{if } \boldsymbol{{L}}_{A} = 1, \boldsymbol{{L}}_{V} = 0 \ ( \textcolor{Magenta}{\textup{Artery}}) \\
min(\boldsymbol{{Y}}_{V}, \boldsymbol{{Y}}_{BV}), & \text{if } \boldsymbol{{L}}_{A} = 0, \boldsymbol{{L}}_{V} = 1 \ (\textcolor{cyan}{\textup{Vein}})\\
min(\boldsymbol{{Y}}_{A}, \boldsymbol{{Y}}_{V}, \boldsymbol{{Y}}_{BV}), & \text{if } \boldsymbol{{L}}_{A} = 1, \boldsymbol{{L}}_{V}=1 \ (\textcolor{gray}{\textup{Crossing}}) \\
\boldsymbol{{Y}}_{BV}, & \text{if } \boldsymbol{{L}}_{A} = 0, \boldsymbol{{L}}_{V}=0 \ (\textcolor{blue}{\textup{Uncertain}} / \textcolor{black}{\textup{BG}}) 
\end{cases} \\
\end{aligned}$}
\label{CCC}
\end{equation}
where $\boldsymbol{L}_A$ and $\boldsymbol{L}_V \in [0,1]$ is the label of artery and vein, respectively, BG means background. According to equation \ref{CCC}, our modified prediction map $\boldsymbol{Y}_{C^3}$ is fused with specialized knowledge about different classes, including Artery, Vein, Crossing, Uncertain Blood Vessel and Background region. With this novel adapted prediction map, we can get our proposed $\mathcal{L}_{C^3}$ as follows: 
\begin{equation}
\begin{aligned}
\mathcal{L}_{C^3}&=\mathcal{L}_{BCE}(\boldsymbol{{Y}}_{C^3}, \boldsymbol{{L}}_{BV})\\
\mathcal{L}_{BCE}(\boldsymbol{Y}, \boldsymbol{L})&=-[{\boldsymbol{L}} \log \boldsymbol{Y}+\left(1-{\boldsymbol{L}}\right) \log \left(1-\boldsymbol{Y}\right)]
\end{aligned}
\end{equation}
where $\boldsymbol{L}_{BV}\in[0,1]$ is the label of blood vessel. 
Next, we analyze the effect of our proposed Channel-Coupled Vessel Consistency loss ($\mathcal{L}_{C^3}$) on A/V classification. From the equation \ref{CCC}, we can conclude that: 

\noindent{\textbf{(1) Semantic Consistency Across Channels:}} Rather than treating the artery, vein and vessel segmentation as three isolated tasks, our proposed $\mathcal{L}_{C^3}$ enforces semantic consistency by integrating their predictions through anatomically grounded rules. This coupling ensures predictions across channels are semantically coherent. For instance, if a pixel is classified as an artery, it must also be recognized as a vessel. Such constraints are implemented by taking the minimum value between the artery and vessel probability maps, thereby eliminating contradictions -- such as a pixel being labeled as an artery but not as part of a vessel.

\noindent{\textbf{(2) Enhanced Robustness in Complex Scenarios:}} Retinal images often present challenges such as artery-vein crossings and ambiguous regions. Our $\mathcal{L}_{C^3}$ explicitly addresses these complexities: (i) Crossing regions (where both artery and vein labels are present, {\itshape i.e.}, $\boldsymbol{L}_{A}=1, \boldsymbol{L}_{V}=1$) are modeled by incorporating the predictions from all three segmentation maps. (ii) Uncertain or background regions (where $\boldsymbol{L}_{A}=0, \boldsymbol{L}_{V}=0$) default to the general vessel prediction, without enforcing a specific artery/vein classification. This targeted treatment enhances the model's robustness in difficult cases that often hinder conventional independent segmentation approaches.

\noindent{\textbf{(3) Stronger Supervision through Fused Learning:}} The fused prediction map $\boldsymbol{Y}_{C^3}$, which integrates anatomical information from all three channels, serves as a richer supervisory signal during training. When incorporated into the loss function $\mathcal{L}_{C^3}$, it guides the model to learn not only class-specific accuracy but also structurally consistent and anatomically plausible representations.

\begin{figure}[t]
\centering
\includegraphics[width=0.5\textwidth]{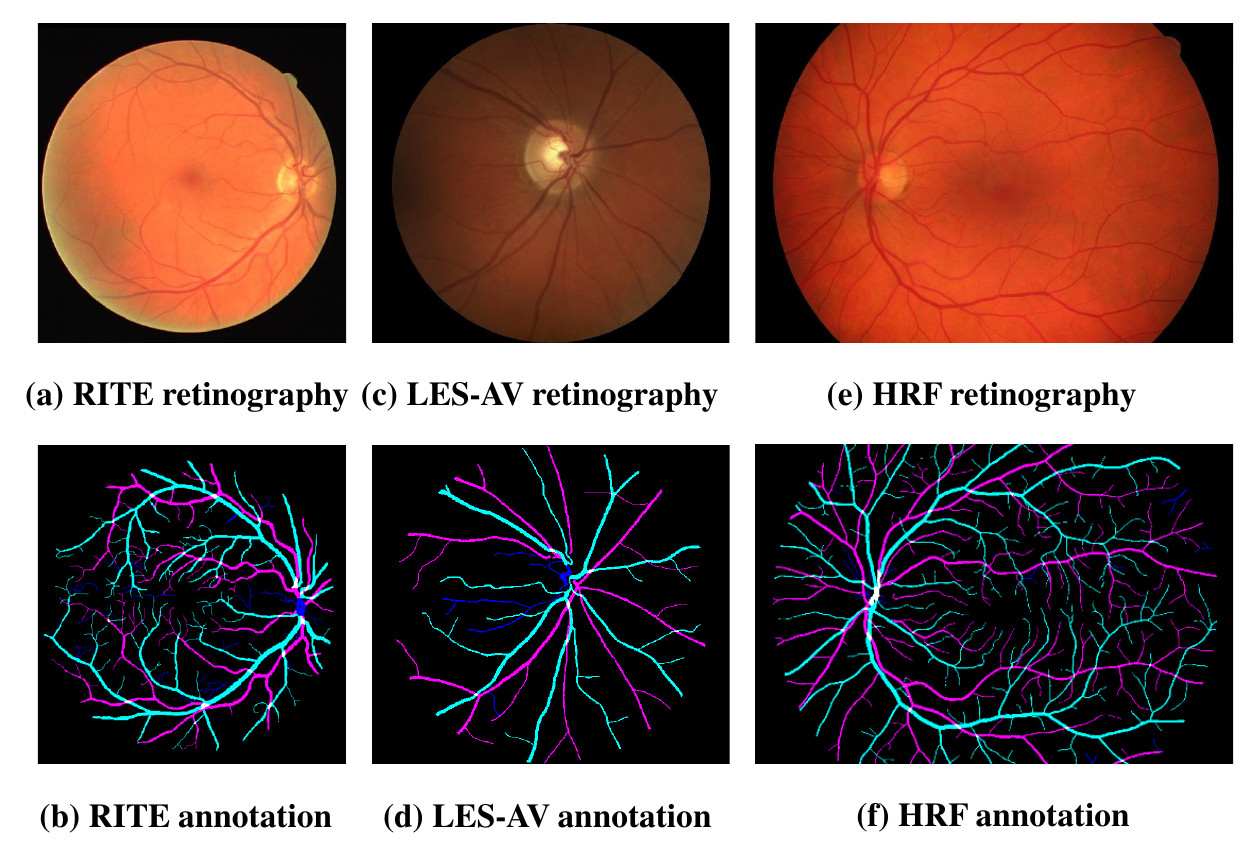}
\caption{Examples of retinal fundus images and their corresponding A/V segmentation maps from different datasets. (a-b) RITE. (c-d) LES-AV. (e-f) HRF. The segmentation maps are visualized as RGB composites, where the red, green and blue channels represent the segmentation masks for \textcolor{Magenta}{arteries}, \textcolor{cyan}{veins} and \textbf{vessels}, respectively. This composition makes arteries appear magenta, veins appear cyan, crossing (regions labeled as both artery and vein) appear white, and uncertain vessels appear blue (because they are identified as vessels but only confidently classified as artery or vein).}
\label{dataset_vis}
\end{figure}

\subsection{Superpixel Guided Contrastive Loss Regularization}\label{sec3.3}

As shown in Figure. \ref{network}(a), following SuperCL \citep{SuperCL}, we use SLIC \citep{SLIC} algorithm to generate the original superpixel mask and then downsample it to guarantee spatial size alignment between the final superpixel mask $\boldsymbol{S}$ and feature map $\boldsymbol{X}^L$ from the encoder. Considering that superpixel can effectively group pixels with similar characteristics within the uniform regions of an image, hence pixels from the same cluster of superpixel mask can be obviously and naturally viewed as positive pairs while pixels from different clusters can be viewed as negative pairs. We can utilize the superpixel mask $\boldsymbol{S}$ to guide contrastive pairs generation for this unsupervised regularization term in Figure. \ref{network}(b). $\mathcal{L}_{intra}$ can be mathematically represented as:

\begin{equation}
\scalebox{0.85}{
$\begin{aligned}
\Omega^{+}: \forall X_{i, j}^L \text{ and } (i, j) &\in S_c, \text{ if } (k, l) \in S_c, \text{ then } \tilde{X}_{k, l}^L \in \Omega^{+}, c \in[1, C] \\
\Omega^{-}: \forall X_{i, j}^L \text{ and } (i, j) &\in S_c, \text{ if } (m, n) \notin S_c, \text{ then } \tilde{X}_{m, n}^L \in \Omega^{-}, c \in[1, C] \\
\mathcal{L}_{intra}&=-\log \frac{\exp \left(\boldsymbol{X}_{\Omega^{+}}\right)}{\exp \left(\boldsymbol{X}_{\Omega^{+}}\right)+\exp \left(\boldsymbol{X}_{\Omega^{-}}\right)}
\end{aligned}$}
\label{SUC}
\end{equation}
$\boldsymbol{S}_c$ denotes the $c_{th}$ cluster of the superpixel map $\boldsymbol{S}$, $C$ denotes the total number of 
superpixel clusters. $|\Omega^+|$ and $|\Omega^-|$ are respectively the set of positive and negative pixel samples for the anchor pixel $(i, j)$. 
As shown in Figure. \ref{network}(b), for an anchor pixel $(i, j)$, $(k, l)$ is its positive pair because $(i, j)$ and $(k, l)$ are in the same superpixel cluster $\boldsymbol{S}_c$; $(m, n)$ is its negative pair because $(i, j)$ and $(m, n)$ are in different superpixel clusters. 
With the introduced $\mathcal{L}_{intra}$, the network can be optimized to extract more discriminative feature-level fine-grained representations with less pixel-level false negative pairs, hence guiding more precise retinal A/V classification.

\begin{table}[t]
\centering
\caption{Proportional distribution (in percentage) of samples (pixels) among various classes across different datasets, as used in training and evaluation.}\label{dataset}
\setlength{\tabcolsep}{14.0pt}
\begin{tabular}{lccc}
\Xhline{1.0pt} 
\multirow{2}{*}{Class} & \multicolumn{3}{c}{Datasets} \\
\cmidrule(lr){2-4}
 & RITE & LES-AV & HRF \\
\Xhline{1.0pt}
{Background} & 87.52 & 90.50 & 89.88 \\
Vessel & 12.48 & 9.50 & 10.12 \\
\ - \textcolor{Magenta}{Artery} & 5.19 & 4.28 & 4.49 \\
\ - \textcolor{cyan}{Vein} & 6.37 & 4.81 & 5.19 \\
\ - \textcolor{gray}{Crossing} & 0.32 & 0.14 & 0.26 \\
\ - \textcolor{blue}{Uncertain} & 0.60 & 0.27 & 0.18 \\
\Xhline{1.0pt}
\end{tabular}
\end{table}

\begin{table*}[htbp]
\centering
\caption{Comparison with the SOTA methods in the tasks of A/V classification. The methods marked with * indicate our reproduced results. The best results are in bold.}\label{SOTA}
\setlength{\tabcolsep}{14.0pt}
\begin{tabular}{ccccccc}
\hline
\addlinespace[0.5ex]
\multirow{2}{*}{Dataset} & \multirow{2}{*}{Method} & \multicolumn{3}{c}{A/V classification} & \multicolumn{2}{c}{BV segmentation} \\
\cmidrule(lr){3-5}
\cmidrule(lr){6-7}
& & Sens. & Spec. & Acc. & Acc. & AUROC \\
\hline
\addlinespace[0.5ex]
\multirow{18}{*}{RITE} 
& \cite{girard2019joint} & 86.30 & 86.60 & 86.50 & 95.70 & 97.20 \\ 
& \cite{galdran2019uncertainty} & 89.00 & 90.00 & 89.00 & 93.00 & 95.00 \\
& \cite{ma2019multi} & 93.40 & 95.50 & 94.50 & 95.70 & 98.10 \\
& \cite{hemelings2019artery} & 95.13 & 92.78 & 93.81 & 96.08 & 88.17 \\ 
& \cite{kang2020avnet} & 88.63 & 92.72 & 90.81 & -- & -- \\
& \cite{morano2021simultaneous} & 87.47 & 90.89 & 89.24 & 96.16 & 98.33 \\
& \cite{galdran2022state} & 88.86 & 96.04 & 92.76 & 96.29 & 98.47 \\ 
& \cite{hatamizadeh2022ravir} & 93.10 & 94.31 & 95.13 & -- & -- \\
& \cite{karlsson2022artery} & 95.10 & 96.00 & 95.60 & 95.60 & 98.10 \\
& \cite{AV-casNet}* & 75.83 & 77.83 & 76.97 & 95.63 & 88.17 \\ 
& \cite{chen2022tw} & 95.38 & 97.20 & 96.34 & 95.75 & 96.29 \\
& \cite{chen2022tw}* & 87.11 & 93.27 & 90.55 & 95.64 & 97.24 \\
& \cite{yi2023retinal} & 94.10 & 93.79 & 95.30 & 96.73 & -- \\
& \cite{hu2024semi} & 93.37 & 95.37 & 94.42 & 95.69 & 98.07 \\ 
& \cite{qureshi2013manually} & 95.80 & 96.82 & 96.37 & 94.76 & -- \\
& \cite{morano2024rrwnet} & 95.73 & \textbf{97.38} & 96.66 & 96.29 & \textbf{98.50} \\
& \cite{morano2024rrwnet}* & 95.03 & 96.75 & 95.99 & 96.20 & 98.49 \\
& \textbf{Ours}& \textbf{96.21} & 97.20 & \textbf{96.77} & \textbf{96.30}  & 98.36 \\
\hline
\addlinespace[0.5ex]
\multirow{8}{*}{LES-AV}
& \cite{galdran2019uncertainty} & 88.00 & 85.00 & 86.00 & - & - \\
& \cite{kang2020avnet} & 94.26 & 90.90 & 92.19 & - & - \\
& \cite{galdran2022state} & 86.86 & 93.56 & 90.47 & 95.69 & 96.27 \\
& \cite{morano2024rrwnet} & 94.30 & 95.25 & 94.81 & 95.61 & 97.72 \\
& \cite{morano2024rrwnet}* & 93.38 & 93.56 & 93.48 & 95.95 & 97.43 \\
& \textbf{Ours} & 94.03 & \textbf{96.14} & \textbf{95.18} & \textbf{96.09} & 97.33 \\
\hline
\addlinespace[0.5ex]
\multirow{13}{*}{HRF}
& \cite{galdran2019uncertainty} & 85.00 & 91.00 & 91.00 & -- & -- \\
& \cite{hemelings2019artery} & -- & -- & 96.98 & -- & -- \\
& \cite{AV-casNet}* & 91.26 & 85.13 & 87.80 & 95.55 & 87.55 \\
& \cite{chen2022tw} & 97.06 & 97.29 & 97.19 & 96.59 & 94.66 \\
& \cite{chen2022tw}* & 95.93 & 96.42 & 96.20 & 96.08 & 93.40 \\ 
& \cite{galdran2022state} & 98.10 & 93.17 & 95.35 & \textbf{96.70} & 98.55 \\
& \cite{karlsson2022artery} & 97.07 & 96.53 & 96.77 & 96.17 & 98.42 \\
& \cite{yi2023retinal} & 96.92 & 96.19 & 95.95 & 96.83 & -- \\
&\cite{hu2024semi} & 93.37 & 95.97 & 94.42 & 96.25 & 98.15 \\
& \cite{hemelings2019artery} & 97.46 & 97.05 & 97.23 & 98.48 & -- \\
& \cite{morano2024rrwnet} & 97.98 & 97.72 & 97.83 & 96.60 & \textbf{98.57} \\
& \cite{morano2024rrwnet}* & \textbf{98.22} & 97.64 & 97.90 & 96.24 & 98.16 \\
& \textbf{Ours} & {98.21} & \textbf{98.33} & \textbf{98.28} & 96.40 & 98.35 \\
\hline
\end{tabular}
\end{table*}

\section{EXPERIMENTAL RESULTS}\label{sec4}
\subsection{Datasets}\label{sec4.1}
Experiments are performed on three publicly available datasets containing color retinal images with corresponding A/V annotations: RITE \citep{hu2013automated}, LES-AV \citep{orlando2018towards} and HRF \citep{budai2013robust}. Figure \ref{dataset_vis} illustrates representative color fundus images and their corresponding ground truth segmentation maps from the three datasets. Table \ref{dataset} summarizes the class-wise distribution of pixel samples -- namely background, artery, vein, crossing, and uncertain -- in each dataset. Further details regarding the datasets are provided below.

\textit{RITE dataset:} RITE \citep{hu2013automated} is derived from the DRIVE \citep{staal2004ridge} dataset, which is specifically designed for research on artery/vein (A/V) classification in retinal images. The dataset consists of 40 color fundus images, split into 20 training and 20 testing images. These images originate from 33 healthy patients and 7 patients with Diabetic Retinopathy (DR). They are all centered on the macula and have a resolution of 565 $\times$ 584 pixels and a field of view of 45 degrees, with a circular region of interest (ROI).  

\textit{LES-AV dataset:} LES-AV \citep{orlando2018towards} comprises 22 fundus images, collected from 11 healthy patients and 11 patients with signs of glaucoma. they are captured at a 30-degree field of view (FOV) and a resolution of 1620 $\times$ 1444 pixels, except one taken at a 45-degree FOV with a resolution of 2196 $\times$ 1958 pixels. Since LES-AV does not provide predefined training and testing splits, we follow the previous work \citep{zhou2021learning} and randomly allocate 11 images for training and the remaining 11 images for testing.

\textit{HRF dataset:} HRF \citep{budai2013robust} consists of 45 high-resolution retinal images, each with a resolution of 3504 $\times$ 2336 pixels. The dataset is evenly distributed across three diagnostic categories: 15 images from healthy individuals, 15 from patients with diabetic retinopathy (DR), and 15 from patients with glaucoma.  
HRF dataset primarily included manual annotations for vessel segmentation without explicit artery/vein classification. Recently, \cite{chen2022tw} introduced novel manual annotations to address this limitation. In this work, we primarily utilize the \cite{chen2022tw} annotations for training and testing, following the previous work \citep{morano2024rrwnet} by using the first five images from each category for testing and the remaining images for training.

\subsection{Implementation details}\label{sec4.3}

The model is implemented using the PyTorch framework and trained on an NVIDIA L40S GPU. We use the Adam optimizer \citep{kingma2014adam} with a constant learning rate of $\alpha = 1 \times 10^{-4}$ and exponential decay rates $\beta_1 = 0.9$ and $\beta_2 = 0.999$. Early stopping is applied if the validation loss does not decrease for 200 consecutive epochs. The batch size is set to 1 during training.
Following \citep{morano2024rrwnet}, full-resolution RITE images are used for training, while LES-AV and HRF images are resized to a width of 576 pixels and 1024 pixels, respectively. The datasets are split into 80\% for training and 20\% for validation. All images undergo offline pre-processing, including global contrast enhancement and local intensity normalization, by the following previous work \citep{morano2021simultaneous}. During training, we apply online data augmentation, including color / intensity variations, affine transformations, horizontal flipping, and random cutout. Finally, all predictions generated from the trained model are upsampled to their original resolution for evaluation. 6 metrics including Sensitivity, Specificity, Accuracy, F1 score, mean Intersection over Union (mIoU) and Area Under the Receiver Operating Characteristic curve (AUROC) are used for classification / segmentation performance evaluation. All related experimental settings are kept consistent with those reported in the original RRWNet paper.

\begin{table*}[htbp]
\centering
\caption{Comparison results with different vessel segmentation loss functions (the segmentation backbone is RRWNet).}\label{losses}
\begin{tabular}{cccccccc}
\Xhline{1.0pt}
\addlinespace[0.5ex]
\multirow{2}{*}{Datasets} & \multirow{2}{*}{Loss Functions} & \multicolumn{5}{c}{A/V classification} \\
\cmidrule(lr){3-7}
\multicolumn{2}{l}{} & Sens. & Spec. & Acc. & F1 & mIoU \\
\Xhline{1.0pt}
\addlinespace[0.5ex]
\multirow{8}{*}{RITE} 
& BCE (baseline) & 95.03 & 96.75 & 95.99 & 95.47 & 91.33 \\
& + Connection Sensitive Loss \citep{li2019connection} & 94.97 & 96.61 & 95.88 & 95.35 & 91.12 \\
& + TopoLoss \citep{hu2019topology} & 93.07 & 96.42 & 94.96 & 94.17 & 88.97 \\
& + Flow-based Loss \citep{jena2021self} & 94.46 & 96.49 & 95.59 & 95.01 & 90.49 \\
& + Supervoxel-based Loss \citep{grim2025efficient}  & 95.63 & 96.31 & 96.01 & 95.49 & 91.36 \\
\cmidrule(lr){2-7}
& + $\boldsymbol{C^3}$ (Ours) & 95.62 & \textbf{97.61} & 96.73 & 96.27 & 92.81 \\
& + Intra (Ours) & 95.39 & 97.41 & 96.51 & 96.06 & 92.41 \\
& + $\boldsymbol{C^3}$ + Intra (Ours) & \textbf{96.21} & 97.20 & \textbf{96.77} & \textbf{96.33} & \textbf{92.93} \\
\Xhline{1.0pt}
\addlinespace[0.5ex]
\multirow{8}{*}{LES-AV}
& BCE (baseline) & 94.45 & 91.41 & 92.75 & 91.98 & 85.15 \\
& + Connection Sensitive Loss \citep{li2019connection} & 94.39 & 93.89 & 94.11 & 93.34 & 87.52 \\
& + TopoLoss \citep{hu2019topology} & 93.71 & 92.41 & 92.97 & 92.00 & 85.18 \\
& + Flow-based Loss \citep{jena2021self} & 93.41 & 95.02 & 94.33 & 93.42 & 87.66 \\
& + Supervoxel-based Loss \citep{grim2025efficient} & 90.23 & 92.09 & 91.30 & 89.80 & 81.50 \\
\cmidrule(lr){2-7}
& + $\boldsymbol{C^3}$ (Ours) & 95.10 & 96.39 & 95.82 & 95.22 & 90.88 \\
& + Intra (Ours) & 93.13 & \textbf{96.33} & 94.90 & 94.21 & 89.05 \\
& + $\boldsymbol{C^3}$ + Intra (Ours) & \textbf{95.91} & 96.02 & \textbf{95.97} & \textbf{95.50} & \textbf{91.39} \\
\Xhline{1.0pt}
\addlinespace[0.5ex]
\multirow{8}{*}{HRF}
& BCE (baseline) & 98.22 & 97.64 & 97.90 & 97.67 & 95.45 \\
& + Connection Sensitive Loss \citep{li2019connection} & 98.30 & 97.57 & 97.90 & 97.67 & 95.44 \\
& + TopoLoss \citep{hu2019topology} & 87.66 & 93.11 & 90.72 & 89.24 & 80.57 \\
& + Flow-based Loss \citep{jena2021self} & 98.01 & 98.28 & 98.16 & 97.94 & 95.96 \\
& + Supervoxel-based Loss \citep{grim2025efficient} & 98.05 & 97.91 & 97.97 & 97.72 & 95.55 \\
\cmidrule(lr){2-7}
& + $\boldsymbol{C^3}$ (Ours) & 98.32 & 98.22 & 98.27 & 98.07 & 96.20 \\
& + Intra (Ours) & \textbf{98.32} & 97.98 & 98.13 & 97.92 & 95.92 \\
& + $\boldsymbol{C^3}$ + Intra (Ours) & 98.21 & \textbf{98.33} & \textbf{98.28} & \textbf{98.08} & \textbf{96.23} \\
\Xhline{1.0pt}
\end{tabular}
\end{table*}

\begin{figure*}[htbp]
\centering
\includegraphics[width=\textwidth]{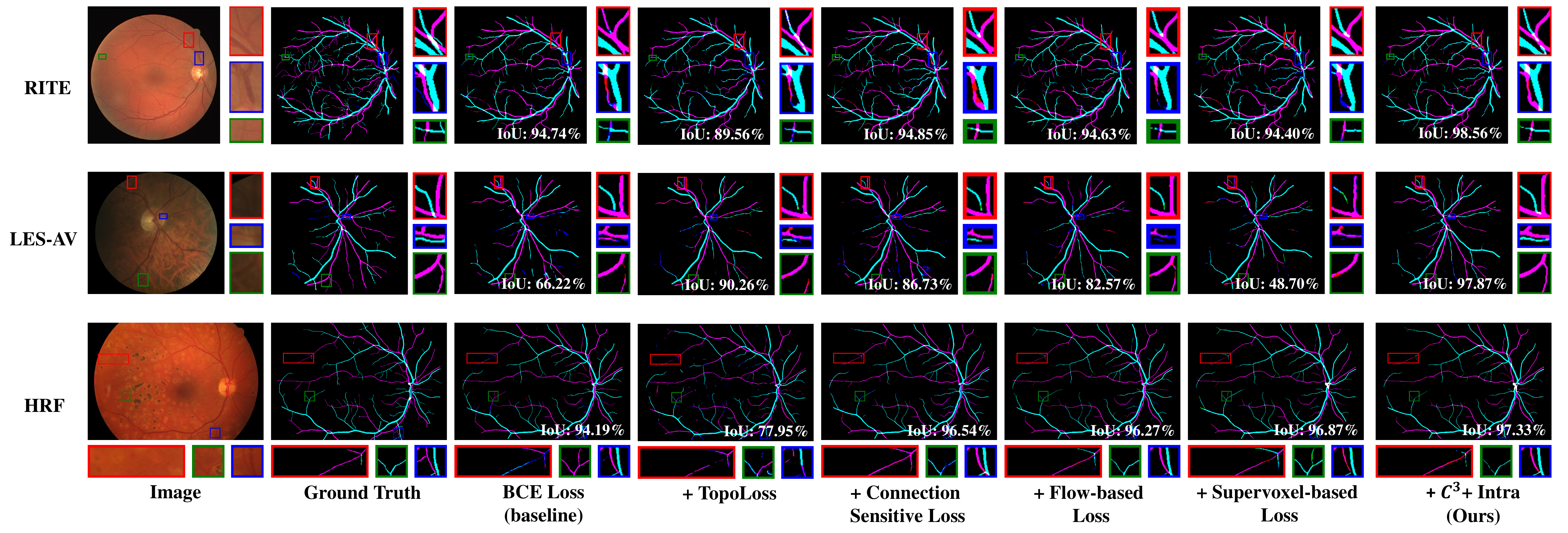}
\caption{Visualization of the comparison results of different vessel segmentation loss functions.}
\label{loss}
\end{figure*}


\subsection{Comparison with SOTA A/V Methods}
Table \ref{SOTA} presents a comparison of A/V classification performance of our proposed $C^3$ loss with Intra loss (network backbone is RRWNet \citep{morano2024rrwnet}) against current state-of-the-art methods for A/V classification and BV segmentation on the RITE, LES-AV and HRF datasets. Notably, our proposed $C^3$ loss and Intra loss are evaluated on the RRWNet backbone. 
According to Table \ref{SOTA}, RRWNet with our losses consistently achieves state-of-the-art A/V classification performance across all the three datasets and most evaluation metrics. Specifically, on RITE, ours achieves an AV classification sensitivity of 96.21\%, accuracy of 96.77\% and BV segmentation accuracy of 96.30\%. And on LES-AV, since the results reported in Table 6 of RRWNet are obtained via cross-dataset evaluation (trained on RITE), hence we use RITE-trained RRWNet optimized with our losses for fair comparison. Ours gets the best classification performance with 96.14\% Spec. and 95.18\% Acc., bringing +0.89\% Spec. and +0.37\% Acc. gain over the second-best method (RRWNet reported results). Finally, on HRF, ours gains an A/V classification specificity of 98.33\% and accuracy of 98.28\%, surpassing the second-best method RRWNet by 0.61\% and 0.38\%, respectively.

\begin{table*}[htbp]
\centering
\caption{Ablation study of our proposed losses for A/V classification on the three A/V datasets using different segmentation backbones (best results in bold).}\label{ablation}
\setlength{\tabcolsep}{4.5pt}
\begin{tabular}{lcccccccccccc}
\Xhline{1.0pt}
\addlinespace[0.5ex]
\multirow{2}{*}{Methods} & \multicolumn{3}{c}{Settings} & \multicolumn{3}{c}{RITE} & \multicolumn{3}{c}{LES-AV} & \multicolumn{3}{c}{HRF} \\
\cmidrule(lr){2-4}
\cmidrule(lr){5-7}
\cmidrule(lr){8-10}
\cmidrule(lr){11-13}
\multicolumn{1}{c}{} & \itshape{$\mathcal{L}_{BCE}$} & \itshape{$\mathcal{L}_{C^3}$} & \itshape{$\mathcal{L}_{intra}$} & Acc. & F1 & mIoU & Acc. & F1 & mIoU & Acc. & F1 & mIoU \\
\Xhline{1.0pt}
\addlinespace[0.5ex]
\multirow{3}{*}{UNet \citep{ronneberger2015u}} 
& \checkmark & & & 93.46 & 92.48 & 86.01 & 91.57 & 90.60 & 82.81 & 96.70 & 96.32 & 92.90 \\
& \checkmark & \checkmark & \multicolumn{1}{c}{} & 94.03 & 93.22 & 87.30 & 91.58 & 90.71 & 83.01 & 96.99 & 96.67 & 93.56  \\
& \checkmark & \checkmark & \checkmark & \textbf{94.32} & \textbf{93.54} & \textbf{87.87} & \textbf{92.08} & \textbf{91.00} & \textbf{83.48} & \textbf{97.10} & \textbf{96.80} & \textbf{93.79} \\

\hline
\addlinespace[0.5ex]
\multirow{3}{*}{IterNet \citep{li2020iternet}} 
& \checkmark & & & 93.70 & 92.88 & 86.71 & 92.37 & 91.62 & 84.53 & 96.23 & 95.80 & 91.94 \\
& \checkmark & \checkmark & & 94.36 & 93.59 & 87.96 & 94.28 & 93.73 & 88.21 & 97.04 & 96.71 & 93.62 \\
& \checkmark & \checkmark & \checkmark & \textbf{94.39} & \textbf{93.64} & \textbf{88.05} & \textbf{94.32} & \textbf{93.77} & \textbf{88.27} & \textbf{97.52} & \textbf{97.23} & \textbf{94.61} \\

\hline
\addlinespace[0.5ex]
\multirow{3}{*}{CTFNet \citep{wang2020ctf}} 
& \checkmark & & & 89.72 & 88.24 & 78.96 & 84.93 & 84.48 & 73.13 & 91.28 & 90.72 & 83.02 \\
& \checkmark & \checkmark & & 90.37 & 88.88 & 79.99 & 86.58 & 85.89 & 75.28 & 93.46 & 92.92 & 86.78 \\
& \checkmark & \checkmark & \checkmark & \textbf{90.79} & \textbf{89.70} & \textbf{81.32} & \textbf{94.11} & \textbf{93.49} & \textbf{87.77} & \textbf{94.05} & \textbf{93.52} & \textbf{87.82} \\

\hline
\addlinespace[0.5ex]
\multirow{3}{*}{AttUNet \citep{oktay2018attention}} 
& \checkmark & & & 93.78 & 92.95 & 86.83 & 92.36 & 91.59 & 84.48 & 97.32 & 97.03 & 94.24  \\
& \checkmark & \checkmark & & 94.75 & 93.99 & 88.66 & 94.35 & 93.73 & 88.20 & 97.77 & 97.53 & 95.18 \\
& \checkmark & \checkmark & \checkmark & \textbf{94.84} & \textbf{94.13} & \textbf{88.90} & \textbf{94.46} & \textbf{93.87} & \textbf{88.46} & \textbf{97.82} & \textbf{97.60} & \textbf{95.30} \\

\hline
\addlinespace[0.5ex]
\multirow{3}{*}{UNet++ \citep{zhou2018unet++}} 
& \checkmark & & & 93.29 & 92.31 & 85.72 & 91.58 & 90.90 & 83.31 & 97.22 & 96.89 & 93.98 \\
& \checkmark & \checkmark & & 93.92 & 93.09 & 87.07 & 93.10 & 92.46 & 85.98 & 97.53 & 97.27 & 94.68 \\
& \checkmark & \checkmark & \checkmark & \textbf{94.11} & \textbf{93.22} & \textbf{87.31} & \textbf{93.59} & \textbf{93.12} & \textbf{87.12}  & \textbf{97.71} & \textbf{97.44} & \textbf{95.01} \\

\hline
\addlinespace[0.5ex]
\multirow{3}{*}{RollingUNet \citep{liu2024rolling}} 
& \checkmark & & & 93.37 & 92.43 & 85.93 & 92.90 & 92.15 & 85.45 & 97.23 & 96.94 & 94.06 \\
& \checkmark & \checkmark & & 94.16 & 93.32 & 87.47 & 93.29 & 92.63 & 86.28 & 97.44 & 97.15 & 94.47 \\
& \checkmark & \checkmark & \checkmark & \textbf{94.56} & \textbf{93.81} & \textbf{88.34} & \textbf{93.49} & \textbf{92.90} & \textbf{86.73} & \textbf{97.60} & \textbf{97.32} & \textbf{94.79} \\
\Xhline{1.0pt}

\addlinespace[0.5ex]
\multirow{3}{*}{RRWNet \citep{RRWNet}}
& \checkmark & & & 95.99 & 95.47 & 91.33 & 92.75 & 91.98 & 85.15 & 97.90 & 97.67 & 95.45 \\
& \checkmark & \checkmark & & 96.73 & 96.27 & 92.81 & 95.82 & 95.22 & 90.88 & 98.27 & 98.07 & 96.20 \\
& \checkmark & \checkmark & \checkmark & \textbf{96.77} & \textbf{96.33} & \textbf{92.93} & \textbf{95.97} & \textbf{95.50} & \textbf{91.39} & \textbf{98.28} & \textbf{98.08} & \textbf{96.23} \\
\Xhline{1.0pt}
\end{tabular}
\end{table*}

\begin{figure*}[htbp]
\centering
\includegraphics[width=\textwidth]{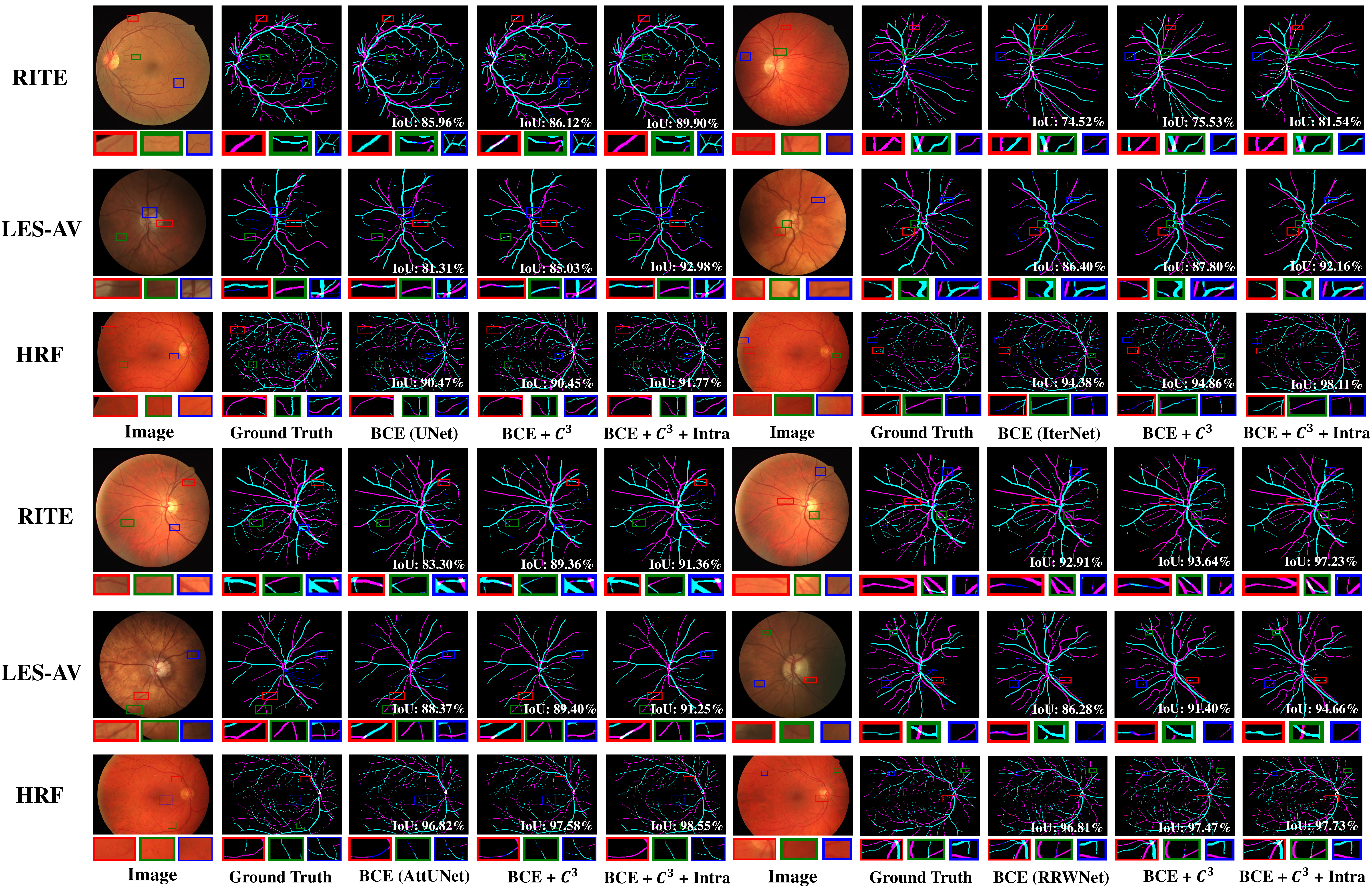}
\caption{Visualization of different segmentation backbones optimized with our proposed $C^3$ and Intra losses on all 3 datasets.}
\label{vis_backbone}
\end{figure*}

\begin{figure*}[t]
\centering
\includegraphics[width=\textwidth]{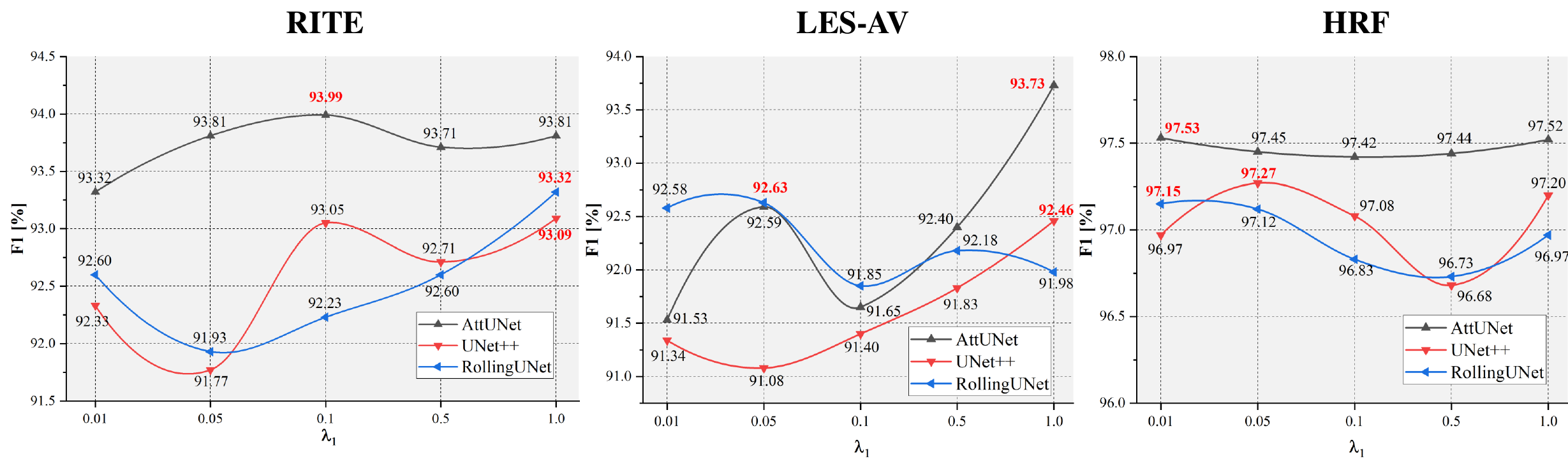}
\caption{Comparison results of different $\lambda_1$, the weighting coefficient of our proposed $C^3$ loss (best results are in red and bold).}
\label{lambda}
\end{figure*}

\begin{figure}[t]
\centering
\includegraphics[width=0.5\textwidth]{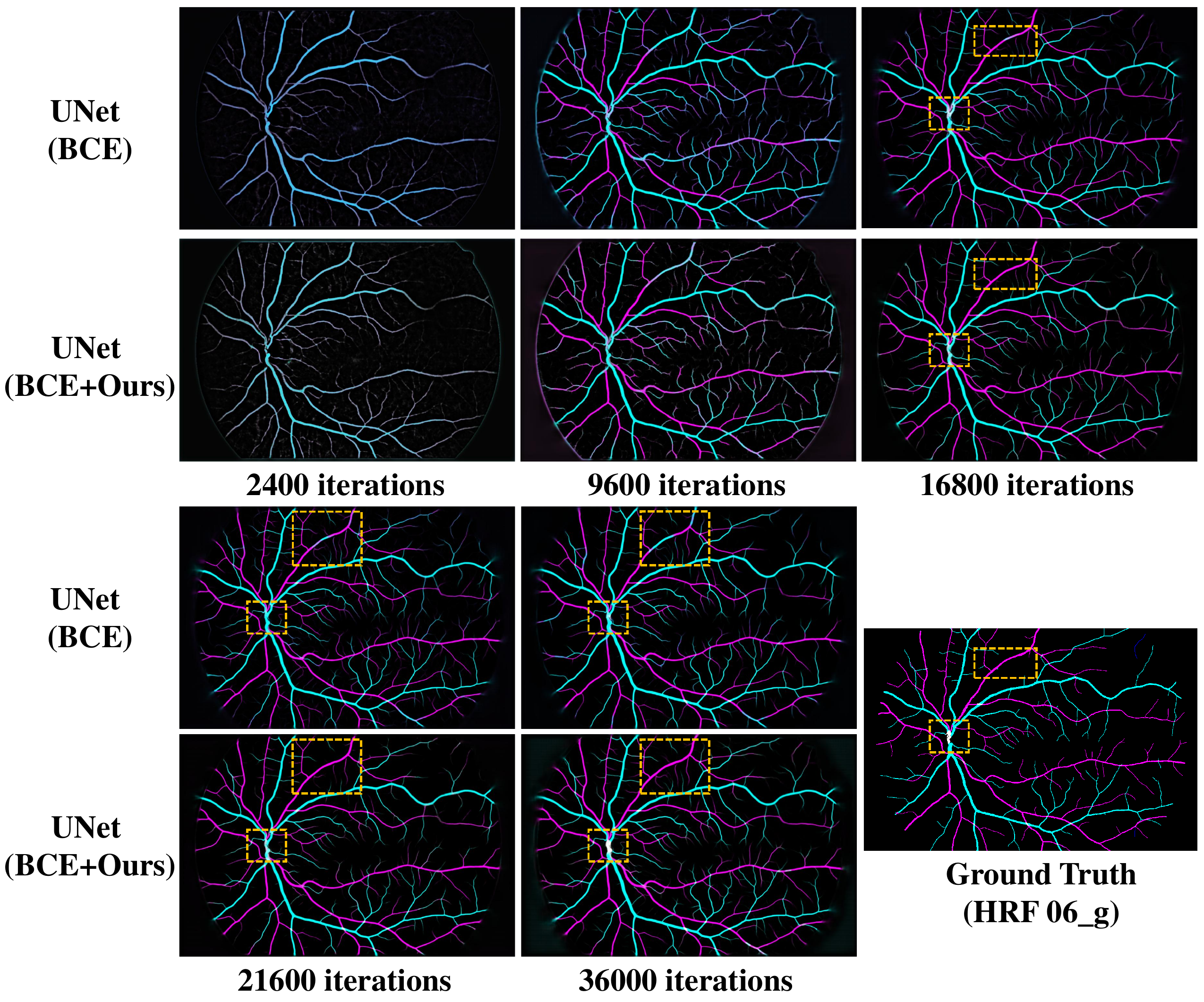}
\caption{Visual comparison of segmentation predictions produced by UNet trained with BCE loss and with our proposed loss at different training iterations (use {\itshape HRF 06\_g} as an example).}
\label{Iteration}
\end{figure}

\begin{figure}[t]
\centering
\includegraphics[width=0.5\textwidth]{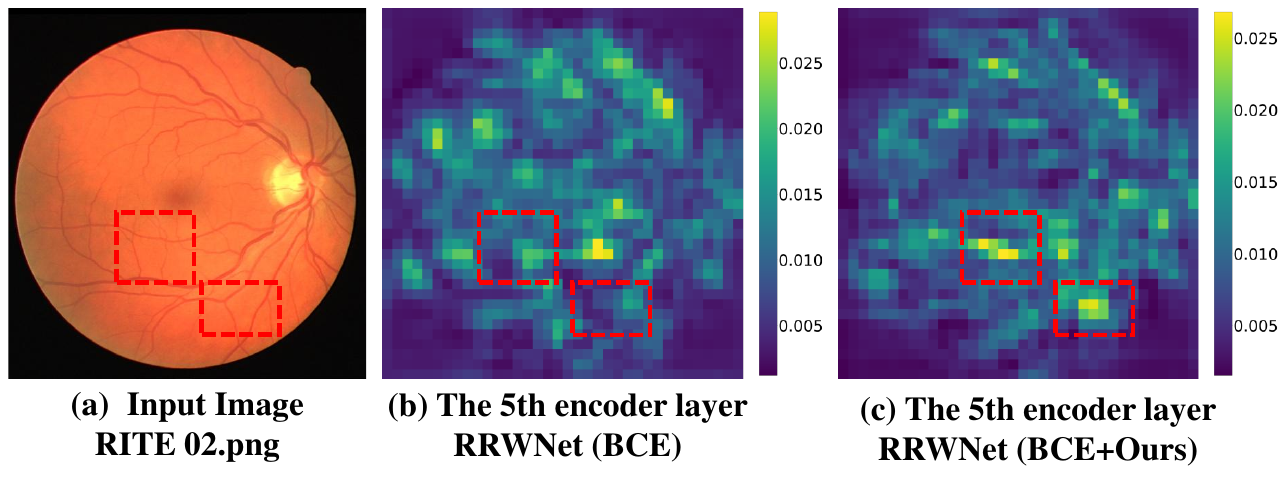}
\caption{The feature maps of the 5th encoder layer of RRWNet optimized with BCE baseline loss and with the addition of our proposed $C^3$ loss, respectively. (use {\itshape RITE 02.png} as an example.)}
\label{feature_map}
\end{figure}

\begin{figure}[t]
  \centering
  {
    \includegraphics[width=0.48\textwidth]{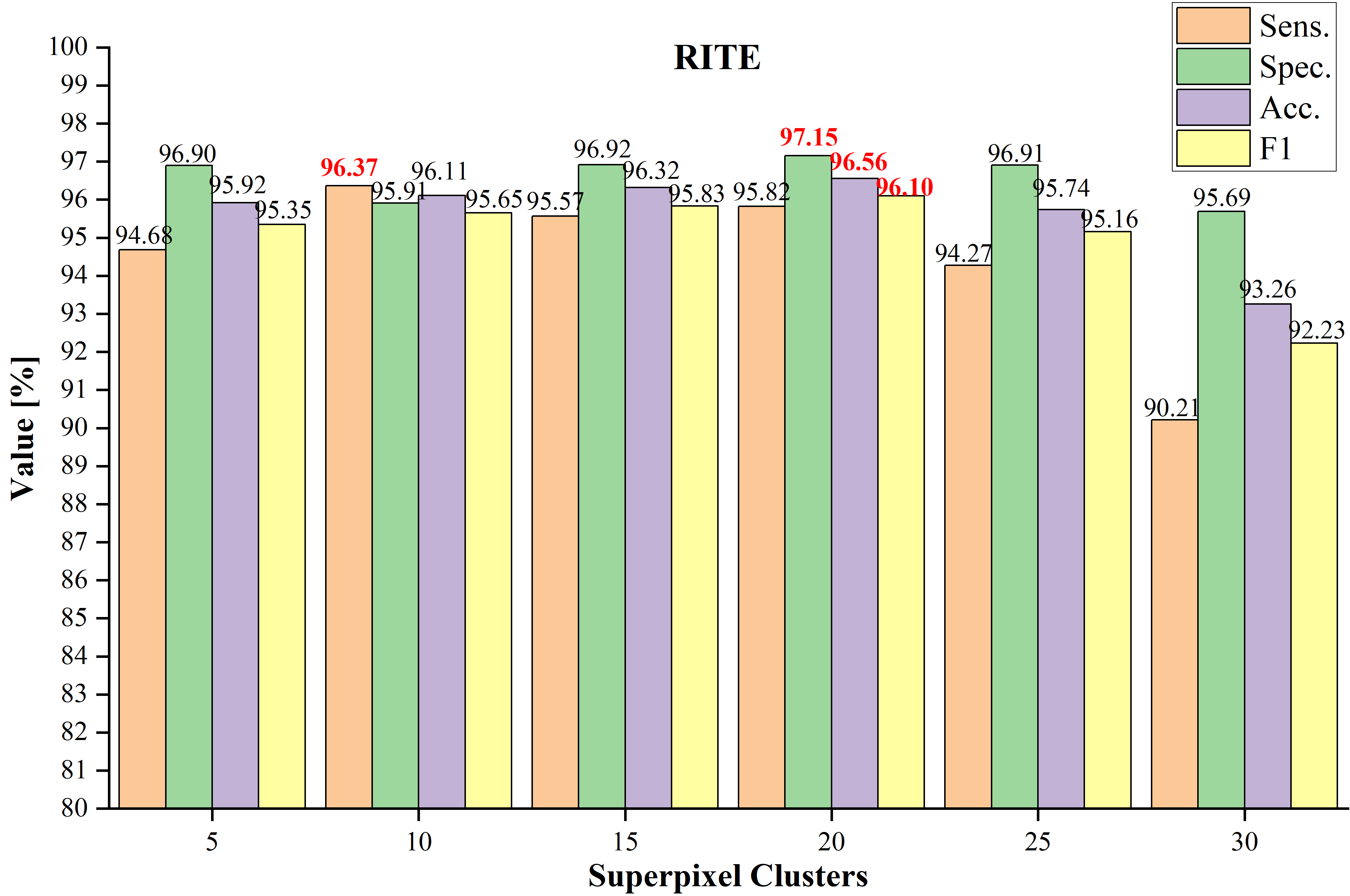}
    \label{fig:image1}
  }
  \vspace{1em} 
  {
    \includegraphics[width=0.48\textwidth]{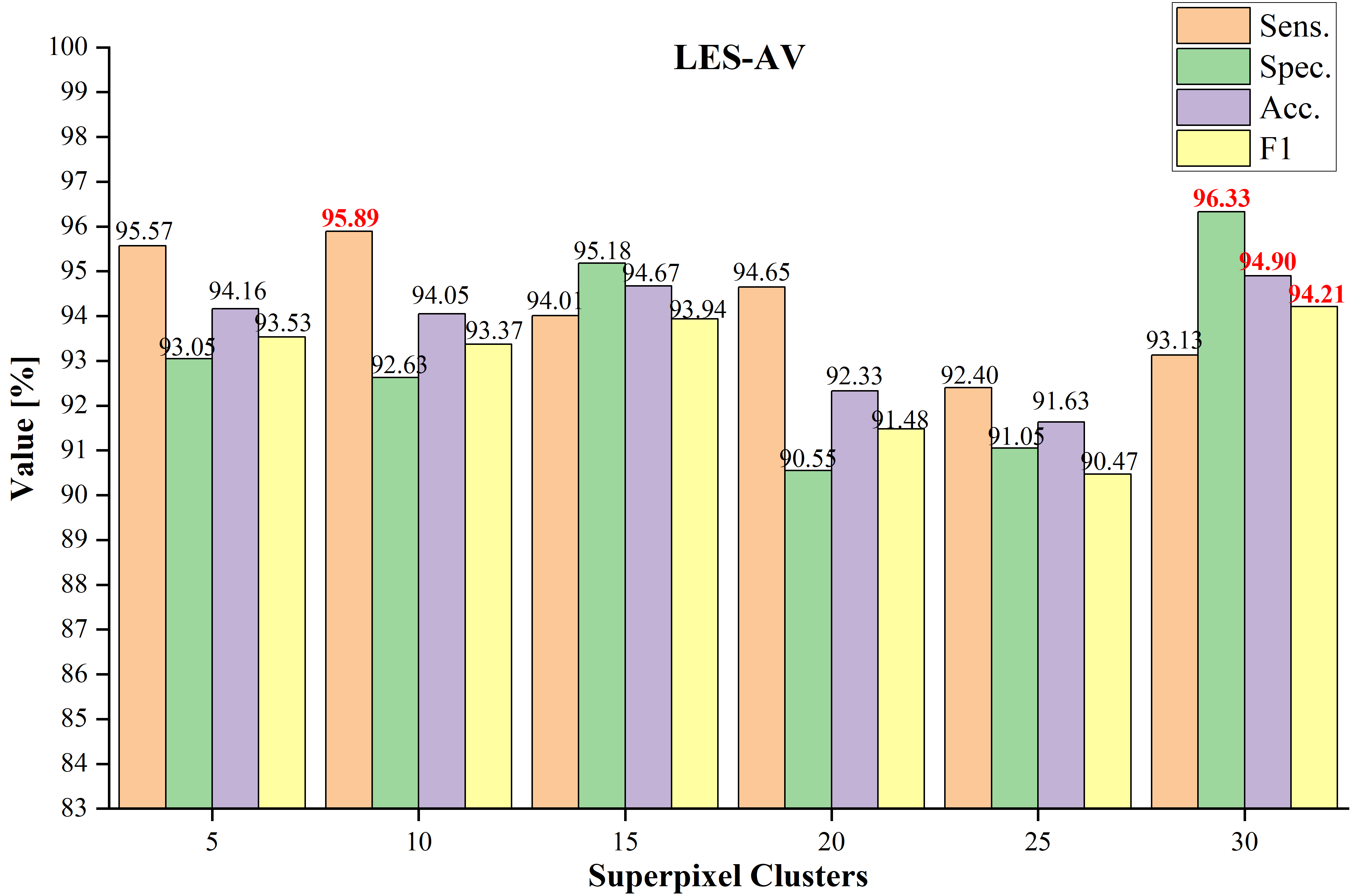}
    \label{fig:image2}
  }
  \vspace{1em}
  {
    \includegraphics[width=0.48\textwidth]{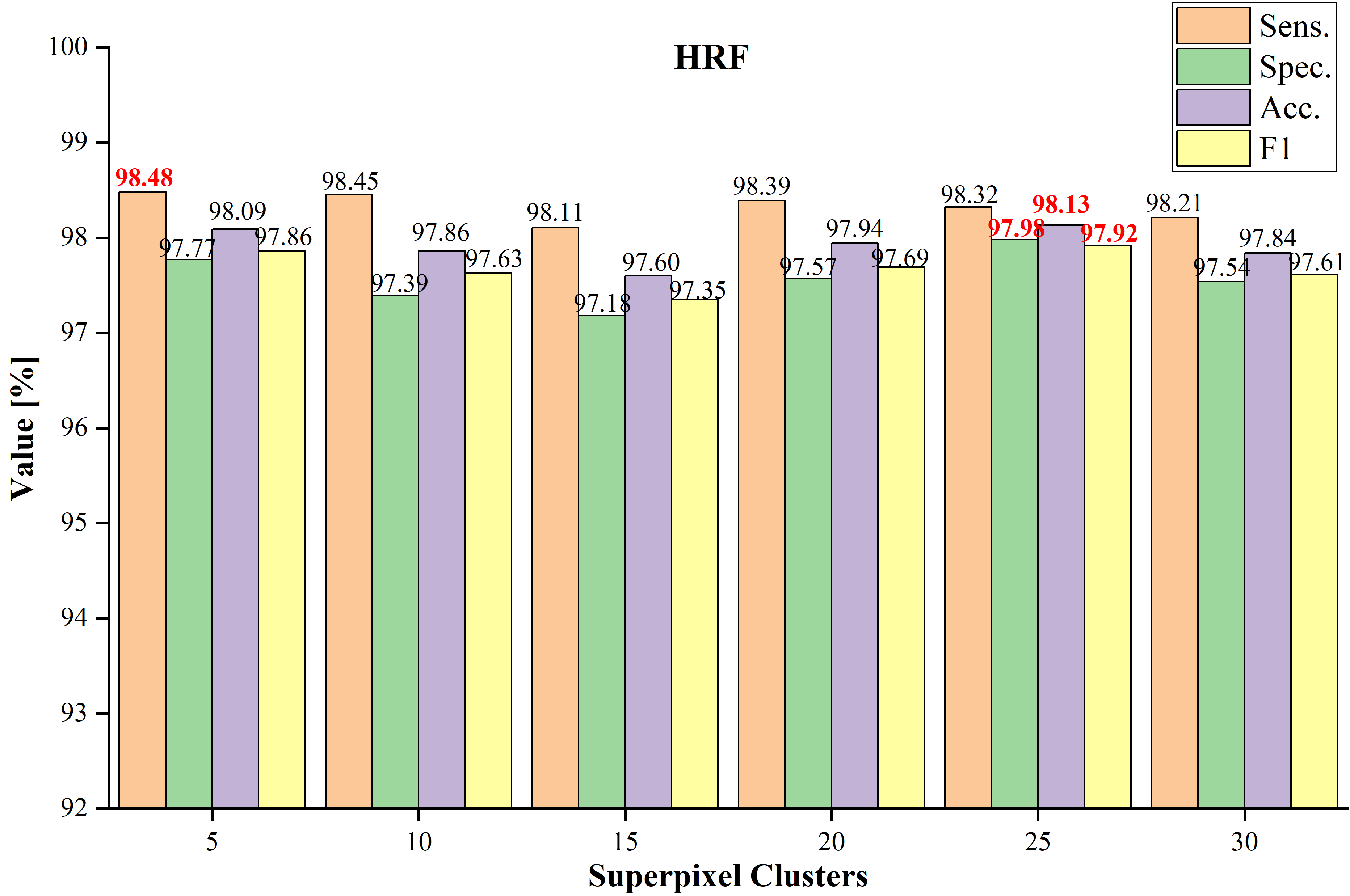}
    \label{fig:image3}
  }
  \caption{Comparison results of different superpixel cluster numbers used in $\mathcal{L}_{intra}$ (best results are in red and bold).}
  \label{cluster}
\end{figure}

\subsection{Comparison with different vessel segmentation loss functions} Additionally, to verify the superiority of our proposed $\mathcal{L}_{C^3}$ and $\mathcal{L}_{intra}$, we further conduct comparison experiments between recently proposed vessel segmentation loss functions with our proposed loss functions. The quantitative results are summarized in Table \ref{losses}, the segmentation backbone is RRWNet with BCE as the baseline loss. According to Table \ref{losses}, our proposed $\mathcal{L}_{C^3}$ and $\mathcal{L}_{intra}$ consistently achieves the best performance across all 3 public datasets and 5 metrics and brings significant improvements over the second-best loss function. Specifically, our BCE + $C^3$ + Intra achieves +0.58\% Sens., +0.76\% Acc., +0.84\% F1, +1.57\% mIoU over Supervoxel-based Loss on RITE; and +1.00\% Spec., 1.64\% Acc., 2.08\% F1, 3.73\% mIoU over Flow-based Loss on LES-AV. We also visualize the results of our proposed loss functions with other vessel segmentation loss functions. As shown in Figure \ref{loss}, RRWNet optimized with the BCE baseline loss and our proposed $C^3$ and Intra losses achieves the best A/V classification performance with 98.56\% IoU on RITE, 97.87\% IoU on LES-AV and 97.33\% on HRF compared with other vessel segmentation loss functions. Moreover, our proposed loss functions also perform well in the classification of tiny micro vessels and vessels of the crossing areas, demonstrating its effectiveness in fine-grained artery-vein classification.

\subsection{Generalization on different segmentation backbones}

We have also conducted ablation experiments to evaluate our proposed $C^3$ and Intra loss on different segmentation backbones, including typical end-to-end retinal vessel segmentation models, like UNet \citep{ronneberger2015u}, IterNet \citep{li2020iternet}, CTFNet \citep{wang2020ctf}, AttUNet \citep{oktay2018attention}, UNet++ \citep{zhou2018unet++}, RollingUNet \citep{liu2024rolling}; and A/V classification model RRWNet \citep{morano2024rrwnet}. The results are summarized in Table \ref{ablation}. According to Table \ref{ablation}, both $\mathcal{L}_{C^3}$ and $\mathcal{L}_{intra}$ can enhance A/V classification performance, and the combination of $\mathcal{L}_{C^3}$ and $\mathcal{L}_{intra}$ (with $\mathcal{L}_{BCE}$ as the baseline loss) achieves the best performance on almost all the metrics across all 3 datasets and 7 different segmentation backbones. {\itshape e.g.}, adding $\mathcal{L}_{C^3}$ results in AttUNet: 1.04\% F1 / 1.83\% mIoU gain, RollingUNet: 0.66\% F1 / 1.14\% mIoU gain and RRWNet: 0.8\% F1 / 1.48\% mIoU gain on RITE. While $\mathcal{L}_{intra}$ brings CTFNet: 7.53\% Acc., 7.6\% F1 and 12.49\% mIoU gain; UNet++: 0.66\% F1 and 1.14\% mIoU gain on LES-AV dataset. Notably, the experimental results on LES-AV in Table \ref{ablation} are obtained by training on the LES-AV training set and testing on the LES-AV test set. The dataset division is described in Section 4.1 Datasets. 
We also visualize the qualitative results of our proposed loss functions on different segmentation backbones across all the 3 datasets. 
According to Figure \ref{vis_backbone}, the application of our proposed $C^3$ and Intra loss significantly enhances A/V classification performance (with a higher IoU) on all the different segmentation backbones, especially in the classification of micro distal vessels and the easily confused vessels in the crossing areas as mentioned in Figure \ref{AV error}. 

\subsection{Comparison results of $\lambda_1$ in $\mathcal{L}_{C^3}$}
According to Table \ref{ablation}, we conclude that our proposed $\mathcal{L}_{C^3}$ matters more than the regularization term $\mathcal{L}_{intra}$, therefore we conduct detailed comparison experiments of the weighting coefficient $\lambda_1$ of $\mathcal{L}_{C^3}$. We validate the optimal value of $\lambda_1$ across different datasets and backbones by selecting from the set [0.01, 0.05, 0.1, 0.5, 1.0]. As shown in Figure \ref{lambda}, on RITE and LES-AV, $\lambda_1=1.0$ yields relatively better performance, {\itshape e.g.}, UNet++ 93.09\% F1, RollingUNet 93.32\% F1 on RITE and AttUNet 93.73\% F1, UNet++ 92.46\% F1 on LES-AV. Whereas on HRF, $\lambda_1=0.01$ proves to be a more suitable choice, {\itshape e.g.}, AttUNet 97.53\% F1 and RollingUNet 97.15\% F1.

\subsection{Visualization analysis of training progress}
Figure \ref{Iteration} shows the segmentation predictions of UNet trained with BCE baseline loss and with the addition of our proposed loss at different training iterations, using {\itshape HRF 06\_g} as an example. According to Figure \ref{Iteration}, we can conclude that: (1) During early training stage (at 2400 and 9600 iterations), the UNet optimized with our proposed loss captures significantly mroe fine-grained micro vessels than the baseline, which is especially evident at 2400 iterations. (2) As training progresses, UNet (BCE + Ours) demonstrates superior performance in challenging regions such as vessel crossings and bifurcations, compared with UNet (BCE). This can be clearly observed in the yellow-boxed areas of images from 16800 to 36000 iterations. On the one hand, our method avoids notable misclassification errors; on the other hand, it effectively distinguishes the crossing regions (white pixel areas), whereas the baseline model (UNet with BCE) tends to misclassify most of the crossing regions as veins. These results suggest that our proposed $C^3$ loss, by incorporating the fused $C^3$ map, provides stronger supervision, thereby effectively enforcing coherence and consistency among vessel, artery, and vein predictions. This leads to better detection of fine vessels in early training stages, and also helps prevent manifest misclassification errors in complex scenarios during later stages of training.   

\subsection{Visualization analysis of encoder feature maps}
Figure \ref{feature_map} illustrates the feature maps of the 5th encoder layer of RRWNet optimized with BCE baseline loss and with the addition of our proposed $C^3$ loss, respectively. According to the red-boxed areas of Figure \ref{feature_map}(b) and (c), RRWNet (BCE + Ours) achieves superior vessel feature extraction compared with RRWNet (BCE), which explains why our method can get more accurate A/V classification performance in the final output prediction map.    

\subsection{Different superpixel cluster numbers in $\mathcal{L}_{intra}$} In $\mathcal{L}_{intra}$, we use SLIC to generate superpixel clusters for contrastive pairs generation. To explore the impact of superpixel numbers on A/V classification performance, we conduct a gradient experiments on the number of superpixel clusters on all 3 datasets. As shown in Figure \ref{cluster}, the results indicate that using 20 superpixel clusters yields the best A/V classification performance on RITE, while 30 and 25 superpixel clusters lead to better results on LES-AV and HRF, respectively. These findings help guide the selection of optimal superpixel configurations for enhancing A/V classification performance across different datasets.

\section{Conclusion}\label{sec6}
In this work, we design a novel loss named Channel-Coupled Vessel Consistency Loss ($\mathcal{L}_{C^3}$) to enforce the coherence and consistency between vessel, artery and vein predictions and avoiding biasing the network toward three simple binary segmentation tasks. Moreover, 
in order to make the network capture more discriminative feature-level fine-grained representations for accurate retinal A/V classification, a regularization term named intra-image pixel-level contrastive loss is introduced by leveraging the structural coherence of superpixels to guide contrastive learning in an unsupervised manner.
Experiments on three A/V classification datasets indicate our proposed $C^3$ loss and Intra loss outperforms existing SOTA A/V classification methods.  

\section*{Acknowledgement}
This work was supported in part by National Natural Science Foundation of China under Grant 82371112, 623B2001, 62394311, in part by Natural Science Foundation of Beijing Municipality under Grant Z210008, and in part by High-grade, Precision and Advanced University Discipline Construction Project of Beijing (BMU2024GJJXK004).


\bibliographystyle{cas-model2-names}

\setcitestyle{numbers}
\bibliography{cas-dc-template}

\begin{thebibliography}{57}
\expandafter\ifx\csname natexlab\endcsname\relax\def\natexlab#1{#1}\fi
\providecommand{\url}[1]{\texttt{#1}}
\providecommand{\href}[2]{#2}
\providecommand{\path}[1]{#1}
\providecommand{\DOIprefix}{doi:}
\providecommand{\ArXivprefix}{arXiv:}
\providecommand{\URLprefix}{URL: }
\providecommand{\Pubmedprefix}{pmid:}
\providecommand{\doi}[1]{\href{http://dx.doi.org/#1}{\path{#1}}}
\providecommand{\Pubmed}[1]{\href{pmid:#1}{\path{#1}}}
\providecommand{\bibinfo}[2]{#2}
\ifx\xfnm\relax \def\xfnm[#1]{\unskip,\space#1}\fi
\bibitem[{Achanta et~al.(2012)Achanta, Shaji, Smith, Lucchi, Fua and
  Süsstrunk}]{SLIC}
\bibinfo{author}{Achanta, R.}, \bibinfo{author}{Shaji, A.},
  \bibinfo{author}{Smith, K.}, \bibinfo{author}{Lucchi, A.},
  \bibinfo{author}{Fua, P.}, \bibinfo{author}{Süsstrunk, S.},
  \bibinfo{year}{2012}.
\newblock \bibinfo{title}{Slic superpixels compared to state-of-the-art
  superpixel methods}.
\newblock \bibinfo{journal}{IEEE Transactions on Pattern Analysis and Machine
  Intelligence} \bibinfo{volume}{34}, \bibinfo{pages}{2274--2282}.
\newblock \DOIprefix\doi{10.1109/TPAMI.2012.120}.
\bibitem[{Achanta and Susstrunk(2017)}]{SNIC}
\bibinfo{author}{Achanta, R.}, \bibinfo{author}{Susstrunk, S.},
  \bibinfo{year}{2017}.
\newblock \bibinfo{title}{Superpixels and polygons using simple non-iterative
  clustering}, in: \bibinfo{booktitle}{Proceedings of the IEEE Conference on
  Computer Vision and Pattern Recognition}.
\bibitem[{Budai et~al.(2013)Budai, Bock, Maier, Hornegger and
  Michelson}]{budai2013robust}
\bibinfo{author}{Budai, A.}, \bibinfo{author}{Bock, R.},
  \bibinfo{author}{Maier, A.}, \bibinfo{author}{Hornegger, J.},
  \bibinfo{author}{Michelson, G.}, \bibinfo{year}{2013}.
\newblock \bibinfo{title}{Robust vessel segmentation in fundus images}.
\newblock \bibinfo{journal}{International journal of biomedical imaging}
  \bibinfo{volume}{2013}, \bibinfo{pages}{154860}.
\bibitem[{Chaitanya et~al.(2020)Chaitanya, Erdil, Karani and Konukoglu}]{GCL}
\bibinfo{author}{Chaitanya, K.}, \bibinfo{author}{Erdil, E.},
  \bibinfo{author}{Karani, N.}, \bibinfo{author}{Konukoglu, E.},
  \bibinfo{year}{2020}.
\newblock \bibinfo{title}{Contrastive learning of global and local features for
  medical image segmentation with limited annotations}.
\newblock \bibinfo{journal}{Advances in Neural Information Processing Systems}
  \bibinfo{volume}{33}, \bibinfo{pages}{12546--12558}.
\bibitem[{Chen et~al.(2020)Chen, Kornblith, Norouzi and
  Hinton}]{chen2020simple}
\bibinfo{author}{Chen, T.}, \bibinfo{author}{Kornblith, S.},
  \bibinfo{author}{Norouzi, M.}, \bibinfo{author}{Hinton, G.},
  \bibinfo{year}{2020}.
\newblock \bibinfo{title}{A simple framework for contrastive learning of visual
  representations}, in: \bibinfo{booktitle}{International conference on machine
  learning}, \bibinfo{organization}{PMLR}. pp. \bibinfo{pages}{1597--1607}.
\bibitem[{Chen et~al.(2022)Chen, Yu, Ma, Ji, Bian, Chu, Shen and
  Zheng}]{chen2022tw}
\bibinfo{author}{Chen, W.}, \bibinfo{author}{Yu, S.}, \bibinfo{author}{Ma, K.},
  \bibinfo{author}{Ji, W.}, \bibinfo{author}{Bian, C.}, \bibinfo{author}{Chu,
  C.}, \bibinfo{author}{Shen, L.}, \bibinfo{author}{Zheng, Y.},
  \bibinfo{year}{2022}.
\newblock \bibinfo{title}{Tw-gan: Topology and width aware gan for retinal
  artery/vein classification}.
\newblock \bibinfo{journal}{Medical Image Analysis} \bibinfo{volume}{77},
  \bibinfo{pages}{102340}.
\bibitem[{Ding et~al.(2014)Ding, Wai, McGeechan, Ikram, Kawasaki, Xie, Klein,
  Klein, Cotch, Wang et~al.}]{ding2014retinal}
\bibinfo{author}{Ding, J.}, \bibinfo{author}{Wai, K.L.},
  \bibinfo{author}{McGeechan, K.}, \bibinfo{author}{Ikram, M.K.},
  \bibinfo{author}{Kawasaki, R.}, \bibinfo{author}{Xie, J.},
  \bibinfo{author}{Klein, R.}, \bibinfo{author}{Klein, B.B.},
  \bibinfo{author}{Cotch, M.F.}, \bibinfo{author}{Wang, J.J.}, et~al.,
  \bibinfo{year}{2014}.
\newblock \bibinfo{title}{Retinal vascular caliber and the development of
  hypertension: a meta-analysis of individual participant data}.
\newblock \bibinfo{journal}{Journal of hypertension} \bibinfo{volume}{32},
  \bibinfo{pages}{207--215}.
\bibitem[{Estrada et~al.(2015)Estrada, Allingham, Mettu, Cousins, Tomasi and
  Farsiu}]{estrada2015retinal}
\bibinfo{author}{Estrada, R.}, \bibinfo{author}{Allingham, M.J.},
  \bibinfo{author}{Mettu, P.S.}, \bibinfo{author}{Cousins, S.W.},
  \bibinfo{author}{Tomasi, C.}, \bibinfo{author}{Farsiu, S.},
  \bibinfo{year}{2015}.
\newblock \bibinfo{title}{Retinal artery-vein classification via topology
  estimation}.
\newblock \bibinfo{journal}{IEEE transactions on medical imaging}
  \bibinfo{volume}{34}, \bibinfo{pages}{2518--2534}.
\bibitem[{Felzenszwalb and Huttenlocher(2004)}]{FH}
\bibinfo{author}{Felzenszwalb, P.F.}, \bibinfo{author}{Huttenlocher, D.P.},
  \bibinfo{year}{2004}.
\newblock \bibinfo{title}{Efficient {Graph-Based} image segmentation}.
\newblock \bibinfo{journal}{Int. J. Comput. Vis.} \bibinfo{volume}{59},
  \bibinfo{pages}{167--181}.
\bibitem[{Galdran et~al.(2022)Galdran, Anjos, Dolz, Chakor, Lombaert and
  Ayed}]{galdran2022state}
\bibinfo{author}{Galdran, A.}, \bibinfo{author}{Anjos, A.},
  \bibinfo{author}{Dolz, J.}, \bibinfo{author}{Chakor, H.},
  \bibinfo{author}{Lombaert, H.}, \bibinfo{author}{Ayed, I.B.},
  \bibinfo{year}{2022}.
\newblock \bibinfo{title}{State-of-the-art retinal vessel segmentation with
  minimalistic models}.
\newblock \bibinfo{journal}{Scientific Reports} \bibinfo{volume}{12},
  \bibinfo{pages}{6174}.
\bibitem[{Galdran et~al.(2019)Galdran, Meyer, Costa, Campilho
  et~al.}]{galdran2019uncertainty}
\bibinfo{author}{Galdran, A.}, \bibinfo{author}{Meyer, M.},
  \bibinfo{author}{Costa, P.}, \bibinfo{author}{Campilho, A.}, et~al.,
  \bibinfo{year}{2019}.
\newblock \bibinfo{title}{Uncertainty-aware artery/vein classification on
  retinal images}, in: \bibinfo{booktitle}{2019 IEEE 16th International
  Symposium on Biomedical Imaging (ISBI 2019)}, \bibinfo{organization}{IEEE}.
  pp. \bibinfo{pages}{556--560}.
\bibitem[{Girard et~al.(2019)Girard, Kavalec and Cheriet}]{girard2019joint}
\bibinfo{author}{Girard, F.}, \bibinfo{author}{Kavalec, C.},
  \bibinfo{author}{Cheriet, F.}, \bibinfo{year}{2019}.
\newblock \bibinfo{title}{Joint segmentation and classification of retinal
  arteries/veins from fundus images}.
\newblock \bibinfo{journal}{Artificial intelligence in medicine}
  \bibinfo{volume}{94}, \bibinfo{pages}{96--109}.
\bibitem[{Grim et~al.(2025)Grim, Chandrashekar and
  S{\"u}mb{\"u}l}]{grim2025efficient}
\bibinfo{author}{Grim, A.}, \bibinfo{author}{Chandrashekar, J.},
  \bibinfo{author}{S{\"u}mb{\"u}l, U.}, \bibinfo{year}{2025}.
\newblock \bibinfo{title}{Efficient connectivity-preserving instance
  segmentation with supervoxel-based loss function}, in:
  \bibinfo{booktitle}{Proceedings of the AAAI Conference on Artificial
  Intelligence}, pp. \bibinfo{pages}{3167--3175}.
\bibitem[{Hatamizadeh et~al.(2022)Hatamizadeh, Hosseini, Patel, Choi, Pole,
  Hoeferlin, Schwartz and Terzopoulos}]{hatamizadeh2022ravir}
\bibinfo{author}{Hatamizadeh, A.}, \bibinfo{author}{Hosseini, H.},
  \bibinfo{author}{Patel, N.}, \bibinfo{author}{Choi, J.},
  \bibinfo{author}{Pole, C.C.}, \bibinfo{author}{Hoeferlin, C.M.},
  \bibinfo{author}{Schwartz, S.D.}, \bibinfo{author}{Terzopoulos, D.},
  \bibinfo{year}{2022}.
\newblock \bibinfo{title}{Ravir: A dataset and methodology for the semantic
  segmentation and quantitative analysis of retinal arteries and veins in
  infrared reflectance imaging}.
\newblock \bibinfo{journal}{IEEE Journal of Biomedical and Health Informatics}
  \bibinfo{volume}{26}, \bibinfo{pages}{3272--3283}.
\bibitem[{He et~al.(2020)He, Fan, Wu, Xie and Girshick}]{he2020momentum}
\bibinfo{author}{He, K.}, \bibinfo{author}{Fan, H.}, \bibinfo{author}{Wu, Y.},
  \bibinfo{author}{Xie, S.}, \bibinfo{author}{Girshick, R.},
  \bibinfo{year}{2020}.
\newblock \bibinfo{title}{Momentum contrast for unsupervised visual
  representation learning}, in: \bibinfo{booktitle}{Proceedings of the IEEE/CVF
  conference on computer vision and pattern recognition}, pp.
  \bibinfo{pages}{9729--9738}.
\bibitem[{Hemelings et~al.(2019)Hemelings, Elen, Stalmans, Van~Keer, De~Boever
  and Blaschko}]{hemelings2019artery}
\bibinfo{author}{Hemelings, R.}, \bibinfo{author}{Elen, B.},
  \bibinfo{author}{Stalmans, I.}, \bibinfo{author}{Van~Keer, K.},
  \bibinfo{author}{De~Boever, P.}, \bibinfo{author}{Blaschko, M.B.},
  \bibinfo{year}{2019}.
\newblock \bibinfo{title}{Artery--vein segmentation in fundus images using a
  fully convolutional network}.
\newblock \bibinfo{journal}{Computerized Medical Imaging and Graphics}
  \bibinfo{volume}{76}, \bibinfo{pages}{101636}.
\bibitem[{Hu et~al.(2024)Hu, Qiu, Wang and Zhang}]{hu2024semi}
\bibinfo{author}{Hu, J.}, \bibinfo{author}{Qiu, L.}, \bibinfo{author}{Wang,
  H.}, \bibinfo{author}{Zhang, J.}, \bibinfo{year}{2024}.
\newblock \bibinfo{title}{Semi-supervised point consistency network for retinal
  artery/vein classification}.
\newblock \bibinfo{journal}{Computers in Biology and Medicine}
  \bibinfo{volume}{168}, \bibinfo{pages}{107633}.
\bibitem[{Hu et~al.(2013)Hu, Abr{\`a}moff and Garvin}]{hu2013automated}
\bibinfo{author}{Hu, Q.}, \bibinfo{author}{Abr{\`a}moff, M.D.},
  \bibinfo{author}{Garvin, M.K.}, \bibinfo{year}{2013}.
\newblock \bibinfo{title}{Automated separation of binary overlapping trees in
  low-contrast color retinal images}, in: \bibinfo{booktitle}{Medical Image
  Computing and Computer-Assisted Intervention--MICCAI 2013: 16th International
  Conference, Nagoya, Japan, September 22-26, 2013, Proceedings, Part II 16},
  \bibinfo{organization}{Springer}. pp. \bibinfo{pages}{436--443}.
\bibitem[{Hu et~al.(2019)Hu, Li, Samaras and Chen}]{hu2019topology}
\bibinfo{author}{Hu, X.}, \bibinfo{author}{Li, F.}, \bibinfo{author}{Samaras,
  D.}, \bibinfo{author}{Chen, C.}, \bibinfo{year}{2019}.
\newblock \bibinfo{title}{Topology-preserving deep image segmentation}.
\newblock \bibinfo{journal}{Advances in neural information processing systems}
  \bibinfo{volume}{32}.
\bibitem[{Jampani et~al.(2018)Jampani, Sun, Liu, Yang and Kautz}]{SSN}
\bibinfo{author}{Jampani, V.}, \bibinfo{author}{Sun, D.}, \bibinfo{author}{Liu,
  M.Y.}, \bibinfo{author}{Yang, M.H.}, \bibinfo{author}{Kautz, J.},
  \bibinfo{year}{2018}.
\newblock \bibinfo{title}{Superpixel sampling networks}, in:
  \bibinfo{booktitle}{Proceedings of the European Conference on Computer
  Vision}.
\bibitem[{Jena et~al.(2021)Jena, Singla and Batmanghelich}]{jena2021self}
\bibinfo{author}{Jena, R.}, \bibinfo{author}{Singla, S.},
  \bibinfo{author}{Batmanghelich, K.}, \bibinfo{year}{2021}.
\newblock \bibinfo{title}{Self-supervised vessel enhancement using flow-based
  consistencies}, in: \bibinfo{booktitle}{Medical Image Computing and Computer
  Assisted Intervention--MICCAI 2021: 24th International Conference,
  Strasbourg, France, September 27--October 1, 2021, Proceedings, Part II 24},
  \bibinfo{organization}{Springer}. pp. \bibinfo{pages}{242--251}.
\bibitem[{Kang et~al.(2020)Kang, Gao, Guo, Xu, Li and Wang}]{kang2020avnet}
\bibinfo{author}{Kang, H.}, \bibinfo{author}{Gao, Y.}, \bibinfo{author}{Guo,
  S.}, \bibinfo{author}{Xu, X.}, \bibinfo{author}{Li, T.},
  \bibinfo{author}{Wang, K.}, \bibinfo{year}{2020}.
\newblock \bibinfo{title}{Avnet: A retinal artery/vein classification network
  with category-attention weighted fusion}.
\newblock \bibinfo{journal}{Computer Methods and Programs in Biomedicine}
  \bibinfo{volume}{195}, \bibinfo{pages}{105629}.
\bibitem[{Karlsson and Hardarson(2022)}]{karlsson2022artery}
\bibinfo{author}{Karlsson, R.A.}, \bibinfo{author}{Hardarson, S.H.},
  \bibinfo{year}{2022}.
\newblock \bibinfo{title}{Artery vein classification in fundus images using
  serially connected u-nets}.
\newblock \bibinfo{journal}{Computer Methods and Programs in Biomedicine}
  \bibinfo{volume}{216}, \bibinfo{pages}{106650}.
\bibitem[{Kingma and Ba(2014)}]{kingma2014adam}
\bibinfo{author}{Kingma, D.P.}, \bibinfo{author}{Ba, J.}, \bibinfo{year}{2014}.
\newblock \bibinfo{title}{Adam: A method for stochastic optimization}.
\newblock \bibinfo{journal}{arXiv preprint arXiv:1412.6980} .
\bibitem[{Li et~al.(2020)Li, Verma, Nakashima, Nagahara and
  Kawasaki}]{li2020iternet}
\bibinfo{author}{Li, L.}, \bibinfo{author}{Verma, M.},
  \bibinfo{author}{Nakashima, Y.}, \bibinfo{author}{Nagahara, H.},
  \bibinfo{author}{Kawasaki, R.}, \bibinfo{year}{2020}.
\newblock \bibinfo{title}{Iternet: Retinal image segmentation utilizing
  structural redundancy in vessel networks}, in:
  \bibinfo{booktitle}{Proceedings of the IEEE/CVF winter conference on
  applications of computer vision}, pp. \bibinfo{pages}{3656--3665}.
\bibitem[{Li et~al.(2019)Li, Li, Li and Zhou}]{li2019connection}
\bibinfo{author}{Li, R.}, \bibinfo{author}{Li, M.}, \bibinfo{author}{Li, J.},
  \bibinfo{author}{Zhou, Y.}, \bibinfo{year}{2019}.
\newblock \bibinfo{title}{Connection sensitive attention u-net for accurate
  retinal vessel segmentation}.
\newblock \bibinfo{journal}{arXiv preprint arXiv:1903.05558} .
\bibitem[{Li and Chen(2015)}]{LSC}
\bibinfo{author}{Li, Z.}, \bibinfo{author}{Chen, J.}, \bibinfo{year}{2015}.
\newblock \bibinfo{title}{Superpixel segmentation using linear spectral
  clustering}, in: \bibinfo{booktitle}{Proceedings of the IEEE Conference on
  Computer Vision and Pattern Recognition}.
\bibitem[{Liu et~al.(2011)Liu, Tuzel, Ramalingam and Chellappa}]{ERS}
\bibinfo{author}{Liu, M.Y.}, \bibinfo{author}{Tuzel, O.},
  \bibinfo{author}{Ramalingam, S.}, \bibinfo{author}{Chellappa, R.},
  \bibinfo{year}{2011}.
\newblock \bibinfo{title}{Entropy rate superpixel segmentation}, in:
  \bibinfo{booktitle}{CVPR 2011}, pp. \bibinfo{pages}{2097--2104}.
\newblock \DOIprefix\doi{10.1109/CVPR.2011.5995323}.
\bibitem[{Liu et~al.(2024)Liu, Zhu, Liu, Yu, Chen and Gao}]{liu2024rolling}
\bibinfo{author}{Liu, Y.}, \bibinfo{author}{Zhu, H.}, \bibinfo{author}{Liu,
  M.}, \bibinfo{author}{Yu, H.}, \bibinfo{author}{Chen, Z.},
  \bibinfo{author}{Gao, J.}, \bibinfo{year}{2024}.
\newblock \bibinfo{title}{Rolling-unet: Revitalizing mlp’s ability to
  efficiently extract long-distance dependencies for medical image
  segmentation}, in: \bibinfo{booktitle}{Proceedings of the AAAI Conference on
  Artificial Intelligence}, pp. \bibinfo{pages}{3819--3827}.
\bibitem[{Long et~al.(2015)Long, Shelhamer and Darrell}]{long2015fully}
\bibinfo{author}{Long, J.}, \bibinfo{author}{Shelhamer, E.},
  \bibinfo{author}{Darrell, T.}, \bibinfo{year}{2015}.
\newblock \bibinfo{title}{Fully convolutional networks for semantic
  segmentation}, in: \bibinfo{booktitle}{Proceedings of the IEEE conference on
  computer vision and pattern recognition}, pp. \bibinfo{pages}{3431--3440}.
\bibitem[{Ma et~al.(2019)Ma, Yu, Ma, Wang, Ding and Zheng}]{ma2019multi}
\bibinfo{author}{Ma, W.}, \bibinfo{author}{Yu, S.}, \bibinfo{author}{Ma, K.},
  \bibinfo{author}{Wang, J.}, \bibinfo{author}{Ding, X.},
  \bibinfo{author}{Zheng, Y.}, \bibinfo{year}{2019}.
\newblock \bibinfo{title}{Multi-task neural networks with spatial activation
  for retinal vessel segmentation and artery/vein classification}, in:
  \bibinfo{booktitle}{Medical Image Computing and Computer Assisted
  Intervention--MICCAI 2019: 22nd International Conference, Shenzhen, China,
  October 13--17, 2019, Proceedings, Part I 22},
  \bibinfo{organization}{Springer}. pp. \bibinfo{pages}{769--778}.
\bibitem[{Mookiah et~al.(2021)Mookiah, Hogg, MacGillivray, Prathiba, Pradeepa,
  Mohan, Anjana, Doney, Palmer and Trucco}]{mookiah2021review}
\bibinfo{author}{Mookiah, M.R.K.}, \bibinfo{author}{Hogg, S.},
  \bibinfo{author}{MacGillivray, T.J.}, \bibinfo{author}{Prathiba, V.},
  \bibinfo{author}{Pradeepa, R.}, \bibinfo{author}{Mohan, V.},
  \bibinfo{author}{Anjana, R.M.}, \bibinfo{author}{Doney, A.S.},
  \bibinfo{author}{Palmer, C.N.}, \bibinfo{author}{Trucco, E.},
  \bibinfo{year}{2021}.
\newblock \bibinfo{title}{A review of machine learning methods for retinal
  blood vessel segmentation and artery/vein classification}.
\newblock \bibinfo{journal}{Medical Image Analysis} \bibinfo{volume}{68},
  \bibinfo{pages}{101905}.
\bibitem[{Morano et~al.(2024a)Morano, Aresta and
  Bogunovi{\'c}}]{morano2024rrwnet}
\bibinfo{author}{Morano, J.}, \bibinfo{author}{Aresta, G.},
  \bibinfo{author}{Bogunovi{\'c}, H.}, \bibinfo{year}{2024}a.
\newblock \bibinfo{title}{Rrwnet: Recursive refinement network for effective
  retinal artery/vein segmentation and classification}.
\newblock \bibinfo{journal}{Expert Systems with Applications}
  \bibinfo{volume}{256}, \bibinfo{pages}{124970}.
\bibitem[{Morano et~al.(2024b)Morano, Aresta and Bogunovi{\'c}}]{RRWNet}
\bibinfo{author}{Morano, J.}, \bibinfo{author}{Aresta, G.},
  \bibinfo{author}{Bogunovi{\'c}, H.}, \bibinfo{year}{2024}b.
\newblock \bibinfo{title}{Rrwnet: Recursive refinement network for effective
  retinal artery/vein segmentation and classification}.
\newblock \bibinfo{journal}{Expert Systems with Applications}
  \bibinfo{volume}{256}, \bibinfo{pages}{124970}.
\bibitem[{Morano et~al.(2021)Morano, Hervella, Novo and
  Rouco}]{morano2021simultaneous}
\bibinfo{author}{Morano, J.}, \bibinfo{author}{Hervella, {\'A}.S.},
  \bibinfo{author}{Novo, J.}, \bibinfo{author}{Rouco, J.},
  \bibinfo{year}{2021}.
\newblock \bibinfo{title}{Simultaneous segmentation and classification of the
  retinal arteries and veins from color fundus images}.
\newblock \bibinfo{journal}{Artificial Intelligence in Medicine}
  \bibinfo{volume}{118}, \bibinfo{pages}{102116}.
\bibitem[{Oktay et~al.(2018)Oktay, Schlemper, Folgoc, Lee, Heinrich, Misawa,
  Mori, McDonagh, Hammerla, Kainz et~al.}]{oktay2018attention}
\bibinfo{author}{Oktay, O.}, \bibinfo{author}{Schlemper, J.},
  \bibinfo{author}{Folgoc, L.L.}, \bibinfo{author}{Lee, M.},
  \bibinfo{author}{Heinrich, M.}, \bibinfo{author}{Misawa, K.},
  \bibinfo{author}{Mori, K.}, \bibinfo{author}{McDonagh, S.},
  \bibinfo{author}{Hammerla, N.Y.}, \bibinfo{author}{Kainz, B.}, et~al.,
  \bibinfo{year}{2018}.
\newblock \bibinfo{title}{Attention u-net: Learning where to look for the
  pancreas}.
\newblock \bibinfo{journal}{arXiv preprint arXiv:1804.03999} .
\bibitem[{Oliveira et~al.(2016)Oliveira, Teixeira, Ren, Cavalcanti and
  Sijbers}]{oliveira2016unsupervised}
\bibinfo{author}{Oliveira, W.S.}, \bibinfo{author}{Teixeira, J.V.},
  \bibinfo{author}{Ren, T.I.}, \bibinfo{author}{Cavalcanti, G.D.},
  \bibinfo{author}{Sijbers, J.}, \bibinfo{year}{2016}.
\newblock \bibinfo{title}{Unsupervised retinal vessel segmentation using
  combined filters}.
\newblock \bibinfo{journal}{PloS one} \bibinfo{volume}{11},
  \bibinfo{pages}{e0149943}.
\bibitem[{Orlando et~al.(2018)Orlando, Barbosa~Breda, Van~Keer, Blaschko,
  Blanco and Bulant}]{orlando2018towards}
\bibinfo{author}{Orlando, J.I.}, \bibinfo{author}{Barbosa~Breda, J.},
  \bibinfo{author}{Van~Keer, K.}, \bibinfo{author}{Blaschko, M.B.},
  \bibinfo{author}{Blanco, P.J.}, \bibinfo{author}{Bulant, C.A.},
  \bibinfo{year}{2018}.
\newblock \bibinfo{title}{Towards a glaucoma risk index based on simulated
  hemodynamics from fundus images}, in: \bibinfo{booktitle}{Medical Image
  Computing and Computer Assisted Intervention--MICCAI 2018: 21st International
  Conference, Granada, Spain, September 16-20, 2018, Proceedings, Part II 11},
  \bibinfo{organization}{Springer}. pp. \bibinfo{pages}{65--73}.
\bibitem[{Qureshi et~al.(2013)Qureshi, Habib, Hunter and
  Al-Diri}]{qureshi2013manually}
\bibinfo{author}{Qureshi, T.A.}, \bibinfo{author}{Habib, M.},
  \bibinfo{author}{Hunter, A.}, \bibinfo{author}{Al-Diri, B.},
  \bibinfo{year}{2013}.
\newblock \bibinfo{title}{A manually-labeled, artery/vein classified benchmark
  for the drive dataset}, in: \bibinfo{booktitle}{Proceedings of the 26th IEEE
  international symposium on computer-based medical systems},
  \bibinfo{organization}{IEEE}. pp. \bibinfo{pages}{485--488}.
\bibitem[{Ronneberger et~al.(2015)Ronneberger, Fischer and
  Brox}]{ronneberger2015u}
\bibinfo{author}{Ronneberger, O.}, \bibinfo{author}{Fischer, P.},
  \bibinfo{author}{Brox, T.}, \bibinfo{year}{2015}.
\newblock \bibinfo{title}{U-net: Convolutional networks for biomedical image
  segmentation}, in: \bibinfo{booktitle}{Medical image computing and
  computer-assisted intervention--MICCAI 2015: 18th international conference,
  Munich, Germany, October 5-9, 2015, proceedings, part III 18},
  \bibinfo{organization}{Springer}. pp. \bibinfo{pages}{234--241}.
\bibitem[{Singh et~al.(2015)Singh, Kumar and Srivastava}]{singh2015local}
\bibinfo{author}{Singh, N.P.}, \bibinfo{author}{Kumar, R.},
  \bibinfo{author}{Srivastava, R.}, \bibinfo{year}{2015}.
\newblock \bibinfo{title}{Local entropy thresholding based fast retinal vessels
  segmentation by modifying matched filter}, in:
  \bibinfo{booktitle}{International Conference on Computing, Communication \&
  Automation}, \bibinfo{organization}{IEEE}. pp. \bibinfo{pages}{1166--1170}.
\bibitem[{Smart et~al.(2015)Smart, Richards, Bhatnagar, Pavesio, Agrawal and
  Jones}]{smart2015study}
\bibinfo{author}{Smart, T.J.}, \bibinfo{author}{Richards, C.J.},
  \bibinfo{author}{Bhatnagar, R.}, \bibinfo{author}{Pavesio, C.},
  \bibinfo{author}{Agrawal, R.}, \bibinfo{author}{Jones, P.H.},
  \bibinfo{year}{2015}.
\newblock \bibinfo{title}{A study of red blood cell deformability in diabetic
  retinopathy using optical tweezers}, in: \bibinfo{booktitle}{Optical trapping
  and optical micromanipulation XII}, \bibinfo{organization}{SPIE}. pp.
  \bibinfo{pages}{342--348}.
\bibitem[{Staal et~al.(2004)Staal, Abr{\`a}moff, Niemeijer, Viergever and
  Van~Ginneken}]{staal2004ridge}
\bibinfo{author}{Staal, J.}, \bibinfo{author}{Abr{\`a}moff, M.D.},
  \bibinfo{author}{Niemeijer, M.}, \bibinfo{author}{Viergever, M.A.},
  \bibinfo{author}{Van~Ginneken, B.}, \bibinfo{year}{2004}.
\newblock \bibinfo{title}{Ridge-based vessel segmentation in color images of
  the retina}.
\newblock \bibinfo{journal}{IEEE transactions on medical imaging}
  \bibinfo{volume}{23}, \bibinfo{pages}{501--509}.
\bibitem[{Tu et~al.(2018)Tu, Liu, Jampani, Sun, Chien, Yang and Kautz}]{SEAL}
\bibinfo{author}{Tu, W.C.}, \bibinfo{author}{Liu, M.Y.},
  \bibinfo{author}{Jampani, V.}, \bibinfo{author}{Sun, D.},
  \bibinfo{author}{Chien, S.Y.}, \bibinfo{author}{Yang, M.H.},
  \bibinfo{author}{Kautz, J.}, \bibinfo{year}{2018}.
\newblock \bibinfo{title}{Learning superpixels with segmentation-aware affinity
  loss}, in: \bibinfo{booktitle}{Proceedings of the IEEE Conference on Computer
  Vision and Pattern Recognition}.
\bibitem[{Wang et~al.(2020)Wang, Zhang, Huang, Wang and Chen}]{wang2020ctf}
\bibinfo{author}{Wang, K.}, \bibinfo{author}{Zhang, X.},
  \bibinfo{author}{Huang, S.}, \bibinfo{author}{Wang, Q.},
  \bibinfo{author}{Chen, F.}, \bibinfo{year}{2020}.
\newblock \bibinfo{title}{Ctf-net: Retinal vessel segmentation via deep
  coarse-to-fine supervision network}, in: \bibinfo{booktitle}{2020 IEEE 17th
  International Symposium on Biomedical Imaging (ISBI)},
  \bibinfo{organization}{IEEE}. pp. \bibinfo{pages}{1237--1241}.
\bibitem[{Welikala et~al.(2017)Welikala, Foster, Whincup, Rudnicka, Owen,
  Strachan, Barman, Eye, Consortium et~al.}]{welikala2017automated}
\bibinfo{author}{Welikala, R.}, \bibinfo{author}{Foster, P.},
  \bibinfo{author}{Whincup, P.}, \bibinfo{author}{Rudnicka, A.R.},
  \bibinfo{author}{Owen, C.G.}, \bibinfo{author}{Strachan, D.P.},
  \bibinfo{author}{Barman, S.}, \bibinfo{author}{Eye, U.B.},
  \bibinfo{author}{Consortium, V.}, et~al., \bibinfo{year}{2017}.
\newblock \bibinfo{title}{Automated arteriole and venule classification using
  deep learning for retinal images from the uk biobank cohort}.
\newblock \bibinfo{journal}{Computers in biology and medicine}
  \bibinfo{volume}{90}, \bibinfo{pages}{23--32}.
\bibitem[{Xu et~al.(2022)Xu, Yang, Wang, Xiao, Xing, Zhang, Wang, Xu, Zhang and
  Lei}]{AV-casNet}
\bibinfo{author}{Xu, X.}, \bibinfo{author}{Yang, P.}, \bibinfo{author}{Wang,
  H.}, \bibinfo{author}{Xiao, Z.}, \bibinfo{author}{Xing, G.},
  \bibinfo{author}{Zhang, X.}, \bibinfo{author}{Wang, W.}, \bibinfo{author}{Xu,
  F.}, \bibinfo{author}{Zhang, J.}, \bibinfo{author}{Lei, J.},
  \bibinfo{year}{2022}.
\newblock \bibinfo{title}{Av-casnet: fully automatic arteriole-venule
  segmentation and differentiation in oct angiography}.
\newblock \bibinfo{journal}{IEEE Transactions on Medical Imaging}
  \bibinfo{volume}{42}, \bibinfo{pages}{481--492}.
\bibitem[{Yang et~al.(2020)Yang, Sun, Jin and Zhou}]{SuperpixelFCN}
\bibinfo{author}{Yang, F.}, \bibinfo{author}{Sun, Q.}, \bibinfo{author}{Jin,
  H.}, \bibinfo{author}{Zhou, Z.}, \bibinfo{year}{2020}.
\newblock \bibinfo{title}{Superpixel segmentation with fully convolutional
  networks}, in: \bibinfo{booktitle}{Proceedings of the CVF Conference on
  Computer Vision and Pattern Recognition}.
\bibitem[{Yi et~al.(2023)Yi, Chen and Yang}]{yi2023retinal}
\bibinfo{author}{Yi, J.}, \bibinfo{author}{Chen, C.}, \bibinfo{author}{Yang,
  G.}, \bibinfo{year}{2023}.
\newblock \bibinfo{title}{Retinal artery/vein classification by multi-channel
  multi-scale fusion network}.
\newblock \bibinfo{journal}{Applied Intelligence} \bibinfo{volume}{53},
  \bibinfo{pages}{26400--26417}.
\bibitem[{Zana and Klein(2001)}]{zana2001segmentation}
\bibinfo{author}{Zana, F.}, \bibinfo{author}{Klein, J.C.},
  \bibinfo{year}{2001}.
\newblock \bibinfo{title}{Segmentation of vessel-like patterns using
  mathematical morphology and curvature evaluation}.
\newblock \bibinfo{journal}{IEEE transactions on image processing}
  \bibinfo{volume}{10}, \bibinfo{pages}{1010--1019}.
\bibitem[{Zeng et~al.(2021)Zeng, Wu, Hu, Xu, Yuan, Huang, Zhuang, Hu and
  Shi}]{PCL}
\bibinfo{author}{Zeng, D.}, \bibinfo{author}{Wu, Y.}, \bibinfo{author}{Hu, X.},
  \bibinfo{author}{Xu, X.}, \bibinfo{author}{Yuan, H.}, \bibinfo{author}{Huang,
  M.}, \bibinfo{author}{Zhuang, J.}, \bibinfo{author}{Hu, J.},
  \bibinfo{author}{Shi, Y.}, \bibinfo{year}{2021}.
\newblock \bibinfo{title}{Positional contrastive learning for volumetric
  medical image segmentation}, in: \bibinfo{booktitle}{International Conference
  on Medical Image Computing and Computer-Assisted Intervention},
  \bibinfo{organization}{Springer}. pp. \bibinfo{pages}{221--230}.
\bibitem[{Zeng et~al.(2025a)Zeng, Lee, Nnamdi, Shi, Tamo, Zhu, He, Zhang, Chen,
  Wang et~al.}]{zeng2025novel}
\bibinfo{author}{Zeng, S.}, \bibinfo{author}{Lee, C.H.},
  \bibinfo{author}{Nnamdi, M.C.}, \bibinfo{author}{Shi, W.},
  \bibinfo{author}{Tamo, J.B.}, \bibinfo{author}{Zhu, L.}, \bibinfo{author}{He,
  H.}, \bibinfo{author}{Zhang, X.}, \bibinfo{author}{Chen, Q.},
  \bibinfo{author}{Wang, M.D.}, et~al., \bibinfo{year}{2025}a.
\newblock \bibinfo{title}{Novel extraction of discriminative fine-grained
  feature to improve retinal vessel segmentation}.
\newblock \bibinfo{journal}{arXiv preprint arXiv:2505.03896} .
\bibitem[{Zeng et~al.(2023)Zeng, Zhu, Zhang, Chen, He, Jin, Tian, Ren, Xie and
  Lu}]{MACL}
\bibinfo{author}{Zeng, S.}, \bibinfo{author}{Zhu, L.}, \bibinfo{author}{Zhang,
  X.}, \bibinfo{author}{Chen, Q.}, \bibinfo{author}{He, H.},
  \bibinfo{author}{Jin, L.}, \bibinfo{author}{Tian, Z.}, \bibinfo{author}{Ren,
  Q.}, \bibinfo{author}{Xie, Z.}, \bibinfo{author}{Lu, Y.},
  \bibinfo{year}{2023}.
\newblock \bibinfo{title}{Multi-level asymmetric contrastive learning for
  volumetric medical image segmentation pre-training}.
\newblock \bibinfo{journal}{arXiv preprint arXiv:2309.11876} .
\bibitem[{Zeng et~al.(2025b)Zeng, Zhu, Zhang, He and Lu}]{SuperCL}
\bibinfo{author}{Zeng, S.}, \bibinfo{author}{Zhu, L.}, \bibinfo{author}{Zhang,
  X.}, \bibinfo{author}{He, H.}, \bibinfo{author}{Lu, Y.},
  \bibinfo{year}{2025}b.
\newblock \bibinfo{title}{Supercl: Superpixel guided contrastive learning for
  medical image segmentation pre-training}.
\newblock \bibinfo{journal}{arXiv preprint arXiv:2504.14737} .
\bibitem[{Zhou et~al.(2021)Zhou, Xu, Hu, Lin, Jacob, Keane and
  Alexander}]{zhou2021learning}
\bibinfo{author}{Zhou, Y.}, \bibinfo{author}{Xu, M.}, \bibinfo{author}{Hu, Y.},
  \bibinfo{author}{Lin, H.}, \bibinfo{author}{Jacob, J.},
  \bibinfo{author}{Keane, P.A.}, \bibinfo{author}{Alexander, D.C.},
  \bibinfo{year}{2021}.
\newblock \bibinfo{title}{Learning to address intra-segment misclassification
  in retinal imaging}, in: \bibinfo{booktitle}{Medical Image Computing and
  Computer Assisted Intervention--MICCAI 2021: 24th International Conference,
  Strasbourg, France, September 27--October 1, 2021, Proceedings, Part I 24},
  \bibinfo{organization}{Springer}. pp. \bibinfo{pages}{482--492}.
\bibitem[{Zhou et~al.(2018)Zhou, Rahman~Siddiquee, Tajbakhsh and
  Liang}]{zhou2018unet++}
\bibinfo{author}{Zhou, Z.}, \bibinfo{author}{Rahman~Siddiquee, M.M.},
  \bibinfo{author}{Tajbakhsh, N.}, \bibinfo{author}{Liang, J.},
  \bibinfo{year}{2018}.
\newblock \bibinfo{title}{Unet++: A nested u-net architecture for medical image
  segmentation}, in: \bibinfo{booktitle}{Deep Learning in Medical Image
  Analysis and Multimodal Learning for Clinical Decision Support: 4th
  International Workshop, DLMIA 2018, and 8th International Workshop, ML-CDS
  2018, Held in Conjunction with MICCAI 2018, Granada, Spain, September 20,
  2018, Proceedings 4}, \bibinfo{organization}{Springer}. pp.
  \bibinfo{pages}{3--11}.
\bibitem[{Zhu et~al.(2021)Zhu, She, Zhang, Lu, Lu, Li and Hu}]{LNSNet}
\bibinfo{author}{Zhu, L.}, \bibinfo{author}{She, Q.}, \bibinfo{author}{Zhang,
  B.}, \bibinfo{author}{Lu, Y.}, \bibinfo{author}{Lu, Z.}, \bibinfo{author}{Li,
  D.}, \bibinfo{author}{Hu, J.}, \bibinfo{year}{2021}.
\newblock \bibinfo{title}{Learning the superpixel in a non-iterative and
  lifelong manner}, in: \bibinfo{booktitle}{Proceedings of the CVF Conference
  on Computer Vision and Pattern Recognition}, pp. \bibinfo{pages}{1225--1234}.

\end{thebibliography}

\end{document}